\documentclass[namedreferences]{SolarPhysics}
\usepackage{graphicx}        
\usepackage{color}           
\usepackage{url}             

\begin{document}
\begin{article}
\begin{opening}

\title{Energy Release During Slow Long Duration Flares Observed by RHESSI}


\author{U. B\c ak-St\c e\' slicka, T. Mrozek and S. Ko\l oma\'nski}

%

%
\institute{Astronomical Institute, University of Wroc{\l}aw, ul. Kopernika 11, 51-622 Wroc{\l}aw, Poland
                     email: \url{bak,mrozek,kolomanski@astro.uni.wroc.pl}}


\begin{abstract}
Slow Long Duration Events (SLDEs) are flares characterized by long duration of rising phase. In many such cases impulsive phase is weak with lack of typical short-lasting pulses. Instead of that smooth, long-lasting Hard X-ray (HXR) emission is observed. We analysed hard X-ray emission and morphology of six selected SLDEs. In our analysis we utilized data from RHESSI and GOES satellites. Physical parameters of HXR sources were obtained from imaging spectroscopy and were used for the energy balance analysis. Characteristic time of heating rate decrease, after reaching its maximum value, is very long, which explains long rising phase of these flares. 
\end{abstract}

%
\end{opening}

\section{Introduction}
A Long Duration Event (LDE) is a flare characterized by slow decrease of Soft X-ray (SXR) emission. SXR images of such flares were observed since {\em Skylab} and {\em Solar Maximum Mission} (SMM) observations (\opencite{sheeley1975}). Those observations indicated that LDEs usually occurred in high arcade of loops (\opencite{kahler1977}). {\em Yohkoh} observed flares with better than {\em Skylab} and SMM temporal and spatial resolution, so that LDEs became the object of study by many authors \cite{tsuneta1992,feldman1995,hudson1996,tomczak1997,harra1998,isobe2002,phillips2005b,kolomanski2007a,kolomanski2007b}.

Among LDEs are flares with rise phase lasting longer than in others, {\em i. e.} more than $30$ minutes. Such flares are called Slow Long Duration Events (SLDEs) and their characteristic feature is that their impulsive phase is weak or does not exist \cite{hudson2000,hudson2001}. In many cases long-lasting Hard X-ray (HXR) emission was observed and its non-thermal character was confirmed (\opencite{hudson2000}). During {\em Yohkoh} mission ($1991-2001$) {\em Geostationary Operational Environmental Satellites} (GOES) registered more than $250$ SLDEs \cite{bak2007}. The well known examples of slow LDE are 1992 February 21 \cite{tsuneta1992} and 1999 January 20 events in which supra-arcade downflows were discovered \cite{mckenzie1999,mckenzie2000}. 

GOES observations were useful for finding SLDEs. In the paper of \inlinecite{bak2005} it has been shown that time interval, $\Delta t$, between the temperature and emission measure maxima, can be used as a measure of duration of rising phase. This method is independent of background level and gives us more reliable value of the duration of rising phase. 

In the work \inlinecite{bak2007} {\em Yohkoh} Hard X-ray Telescope (HXT, \opencite{kosugi1991}) data were used to analyse several limb or near-the-limb slow LDEs. In most of the cases long-lasting HXR emission with lack of typical, short lasting pulses was observed. HXR emission in low energy channel was spatialy correlated with tops of loops seen in SXR images. In a few cases high energy emission sources were observed near footpoints of loops. 

In the paper of \inlinecite{bak2005} {\em Yohkoh} Soft X-ray Telescope (SXT, \opencite{tsuneta1991}) images were used to analyse morphology of loop-top sources (LTS) and to calculate their physical parameters. This analysis indicated that most of the slow LDEs occurred in high or mid-high structures ($h\sim20-50$~Mm). The loop-top sources were characterized by low temperature ($T<10$~MK), low density ($N\sim10^{10}$\,cm$^{–-3}$) and large size ($r>7 \times10^8$ cm). The thermal energy release rate was small, below $1$\,erg\,cm$^{-3}$s$^{-1}$ and decreased very slowly with time after reaching its maximum value.\\
  
Slow LDEs could not be investigated in more detail because of limited sensitivity of SXT telescope and low spectral resolution of HXT telescope. To overcome these instrumental limitations we decided to use {\em Reuven Ramaty High Energy Solar Spectroscopic Imager} (RHESSI) observations. RHESSI allows us to investigate spatially resolved HXR emission of SLDEs with $1\rm keV$ energy resolution, distinguish between the thermal and non-thermal nature of an LTS and calculate its physical parameters. Here we present analysis of six selected slow LDEs.

\section{Observations}
Our analysis is based on RHESSI (\opencite{lin2002}) and GOES X-ray Sensor (XRS) data  supported by {\em Solar and Heliospheric Observatory} Extreme UV Imaging Telescope (SOHO/EIT, \opencite{delaboudiniere1995}), {\em Transition Region and Coronal Explorer} (TRACE, \opencite{handy1999}) and GOES Solar X-Ray Imager (SXI, \opencite{hill2005}) observations. RHESSI is a rotating Fourier imager with nine detectors made of pure germanium crystals (\opencite{lin2002}). The detectors record energy and time of arrival for each HXR photon detected. Pairs of grids placed ahead of the detectors and the rotation of the whole satellite ($4$ s period) cause modulation of HXR radiation coming from solar sources. It is possible to reconstruct HXR images (\opencite{hurford2002}) using several available algorithms. 

We analysed rise phase of SLDE events using images reconstructed with PIXON algorithm (\opencite{puetter1999} and references within). The images were reconstructed in time intervals of $20-40$~s covering the whole rise phase. Energy intervals were narrow, usually $1-2$~keV. Such intervals were chosen since we were interested in the image spectroscopy of observed HXR sources. The image spectroscopy we performed has an advantage in comparison to "standard" spectroscopy based on the fluxes measured for the whole Sun. Namely, for the rotating Fourier imager the background photons do not influence modulation profile, the analysis of fluxes can be done more precisely. 
 
\section{Data Analysis}
For our analysis we selected six limb or near-the-limb solar flares with slow rise phases (the rise phases lasted between $25$--$150$ minutes). We calculated temperature $T_{G}$, emission measure $EM_{G}$ and time difference, $\Delta t$, between their maxima using GOES/XRS data. We obtained physical parameters of loop-top sources using RHESSI data. 

\subsection{Physical Parameters of LTS}
A loop-top source was defined as an area of HXR emission bounded by the intensity isoline  equal to 50\% of intensity of the brightest pixel. We calculated the source area for a given time interval and different energy ranges. The mean area was used for calculating the mean radius, $r$, assuming the spherical shape of the loop-top source. Using the mean radius (Table~\ref{table}) we calculated the volume of emitting region for a given time interval. The altitude of the loop-top source was calculated from centroids of the HXR emission source.

With spatially resolved signals we were able to perform spectroscopy of individual HXR sources (bounded by isocontour equal to 50\% of intensity of brightest pixel) using standard OSPEX package. The best fit to the observed spectra was obtained with a use of thermal, two spectral line complexes (at $6.7$~keV and $8.0$~keV) and single power-law components. Other sets of components, for example double thermal despite thermal plus power-law, were also used. In such cases the obtained fits were significantly worse. The values obtained for $T_{R}$ and $EM_{R}$ were used to calculate heating rate during the rise phase of investigated flares.

Uncertainties in investigated parameters were calculated as follows. In the case of SLDEs we may assume slow temporal changes of the parameters. Thus, observed fast and small fluctuations of the parameters observed for consecutive time intervals may be the product of observational errors. After subtracting a slow-varying trend in: size of the loop-top sources, temperature and emission measure we calculated a standard deviation for them. We treated the standard deviation as an uncertainty of the parameters. The obtained estimations for uncertainties are: about $5\%$ for the mean radius of the loop-top source, about $15\%$ for the volume of LTS, less than $5\%$ for temperature, and less than $20\%$ for emission measure.

A non-thermal component of the HXR spectrum was observed in four cases. We fitted this component with a double power-law function. We treated a break energy and power law index above this energy as free parameters. The power law index below the break energy was fixed at constant value. The total power in non-thermal electrons above cut-off energy, $\epsilon_c$, was calculated using a formula given by \opencite{aschwanden2005}:
\begin{equation}
P\left(\epsilon\geq\epsilon_c\right)=4.3\times10^{24}\frac{b\left(\gamma\right)}{\left(\gamma-1\right)}A\left(\epsilon_c\right)^{-\left(\gamma-1\right)}\ \ \ \ \ \left({\rm erg}\ {\rm s}^{-1}\right)
\end{equation}
Where $b\left(\gamma\right)\approx0.27\gamma^3$ is the auxiliary function calculated in \inlinecite{hudson1978}, 
$A$ is a normalization factor. Since we were interested in a rough estimation of non-thermal power we assumed that break energy from our fits is close to the cut-off energy. 

\subsection{Energy balance}
We estimated the value of the heating function, $E_H$, for the loop-top sources seen in RHESSI images for the all analysed flares. 
In order to calculate heating rate of an LTS we considered its energy balance during the rise phase. Here we consider time interval, $\Delta t$, after temperature maximum, when the energy release has begun to decrease, but the emission measure is still increasing. Three major cooling processes where included into this balance: expansion, radiation and conduction:

\begin{equation}
\label{e4}
\left (\frac{d{\mathcal E}}{dt}\right )_{obs} = \left( \frac{d{\mathcal E}}{dt} \right)_{ad} - E_C - E_R + E_H
\end{equation}

where: 
${\mathcal E} = 3NkT$ is thermal energy density, $\left (\frac{d{\mathcal E}}{dt}\right )_{obs}$ is the decrease of ${\mathcal E}$ per second estimated from temperature ($T$) and density ($N$) values, $\left (\frac{d{\mathcal E}}{dt}\right )_{ad}$ is the decrease due to the adiabatic expansion of plasma in a source, $E_C$ is the energy loss due to thermal conduction, $E_R$ is the radiative loss, and $E_H$ is the heating rate or thermal energy release. The values of $E_C$, $E_R$ and $E_H$ are in erg~cm$^{-3}$~s$^{-1}$. We calculated:

\begin{itemize}
	\item $\left (\frac{d{\mathcal E}}{dt}\right )_{ad} = 5kT\left (\frac{dN}{dt}\right)$,
	\item $E_C = 3.9\times10^{-7}T^{3.5}/(Lr)$, where $r$ is the LTS radius and $L$ is loop semi-length (\opencite{jakimiec1997}). 
	\item $E_R = N^{2}\Phi(T)$, where $\Phi(T)$ is the radiative loss function taken from \inlinecite{dere2009}. 
\end{itemize}

We took the height of an LTS above the photosphere, $h$, as an approximation for $L$ in the expression for $E_C$. Since the value of $h$ is always smaller than $L$ thus, the calculated value of $E_H$ is the upper limit.

Uncertainties calculated for basic parameters allowed us to estimate the errors in the $E_H$. In the case of SLDEs $E_C$ is a dominant contributor in energy balance. It is about an order of magnitude greater than other contributors to the $E_H$. The estimated value of the uncertainty of $E_C$ is about $20\%$. Similar value have been obtained through the analysis of a standard deviation.

For all selected flares we analysed time evolution of $E_{H}(t)$ and calculated characteristic time $\tau$ of the $E_H(t)$ decrease after reaching its maximum value:
\begin{equation}
\label{tau}
\frac{1}{\tau}=\frac{dlnE_{H}(t)}{dt}
\end{equation}


\begin{table}[!hb]
	\caption{Data for selected flares}
	\vspace{3mm}
	\label{table1}
		\begin{tabular}{|c|c|c|c|c|c|c|c|c|c|c|}
\hline

&&\multicolumn{6}{c|}{GOES/XRS}\\
Flare&Date&Start&Max&class&$T_{Gmax}$&$EM_{Gmax}$&$\Delta t$\\
&&[UT]&[UT]&&[MK]&[$10^{48}$cm$^{-3}$]&[min]\\
\hline
1.&2003 Oct. 24&02:19&02:54&M7.6&17.8&43&21\\
2.&2003 Nov. 18&09:23&10:11&M5.4&12.8&29&23\\
3.&2005 Jul. 13&14:01&14:49&M5.0&19.0&26&39\\
4.&2005 Aug. 23&14:19&14:44&M2.7&20.2&14&20\\
5.&2005 Sep. 06&19:32&22:02&M1.4&16.2&7.7&101\\
6.&2007 Jan. 25&06:33&07:14&C6.3&13.0&5&29\\
\hline
\multicolumn{8}{@{} l @{}}{$\Delta t=t(EM_{Gmax})-t(T_{Gmax})$} \\
	\end{tabular}
\end{table}
\section{Results}
We calculated temperature, $T_{G}$, emission measure, $EM_{G}$, and difference between their maxima (Table\,\ref{table1}) using GOES/XRS data. For all flares we used RHESSI data to obtain physical parameters of loop-top sources from spectra fitting. The parameters were used to determine heating rate and characteristic time of its decrease after reaching maximum value. Below we describe three flares in details. Results for all flares are given in Table\,\ref{table}.

\subsection{2003 October 24 flare}

\begin{figure}[t] 
 \includegraphics[width=0.47\textwidth,clip=]{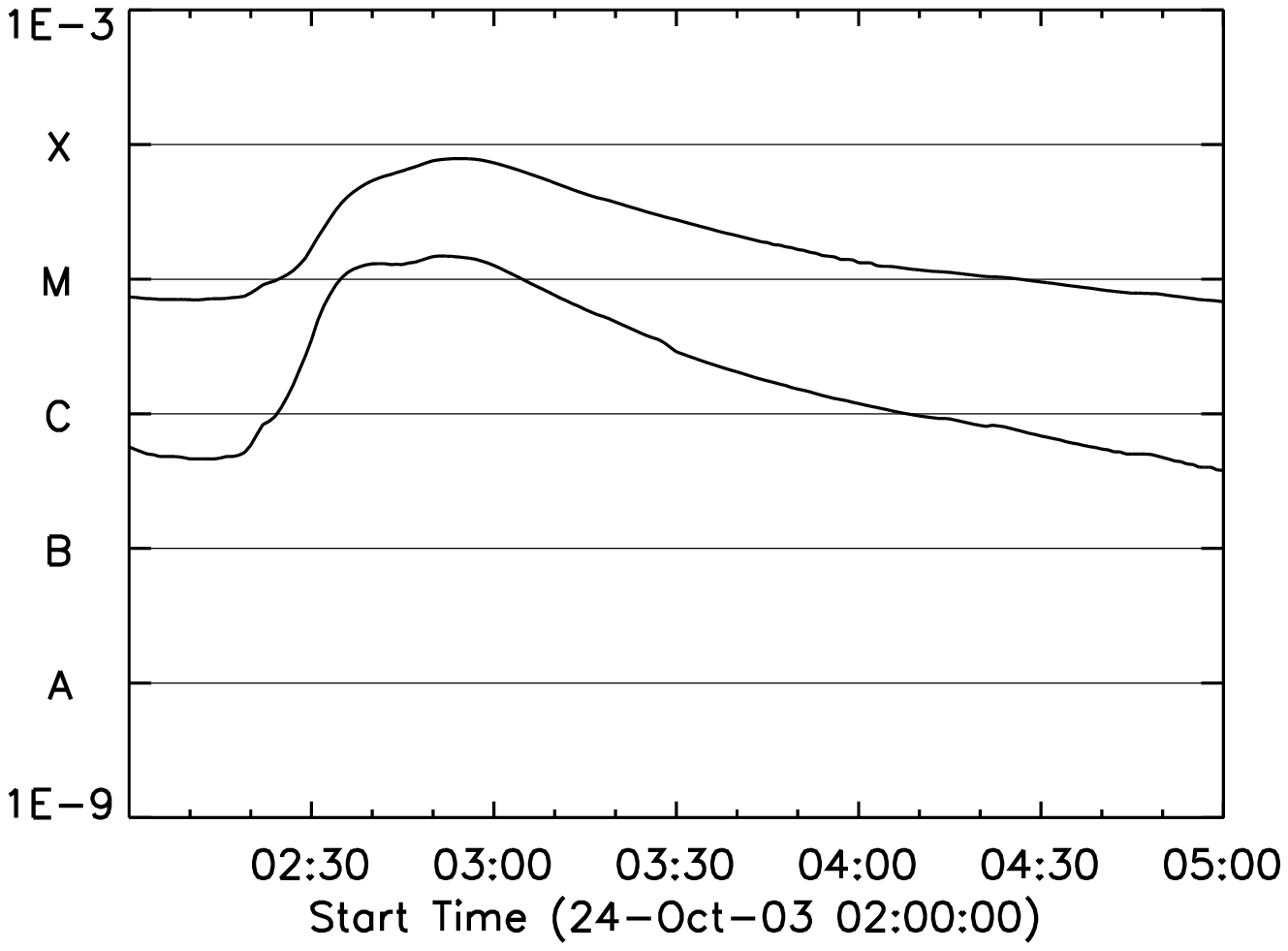}
 \includegraphics[width=0.5\textwidth,clip=]{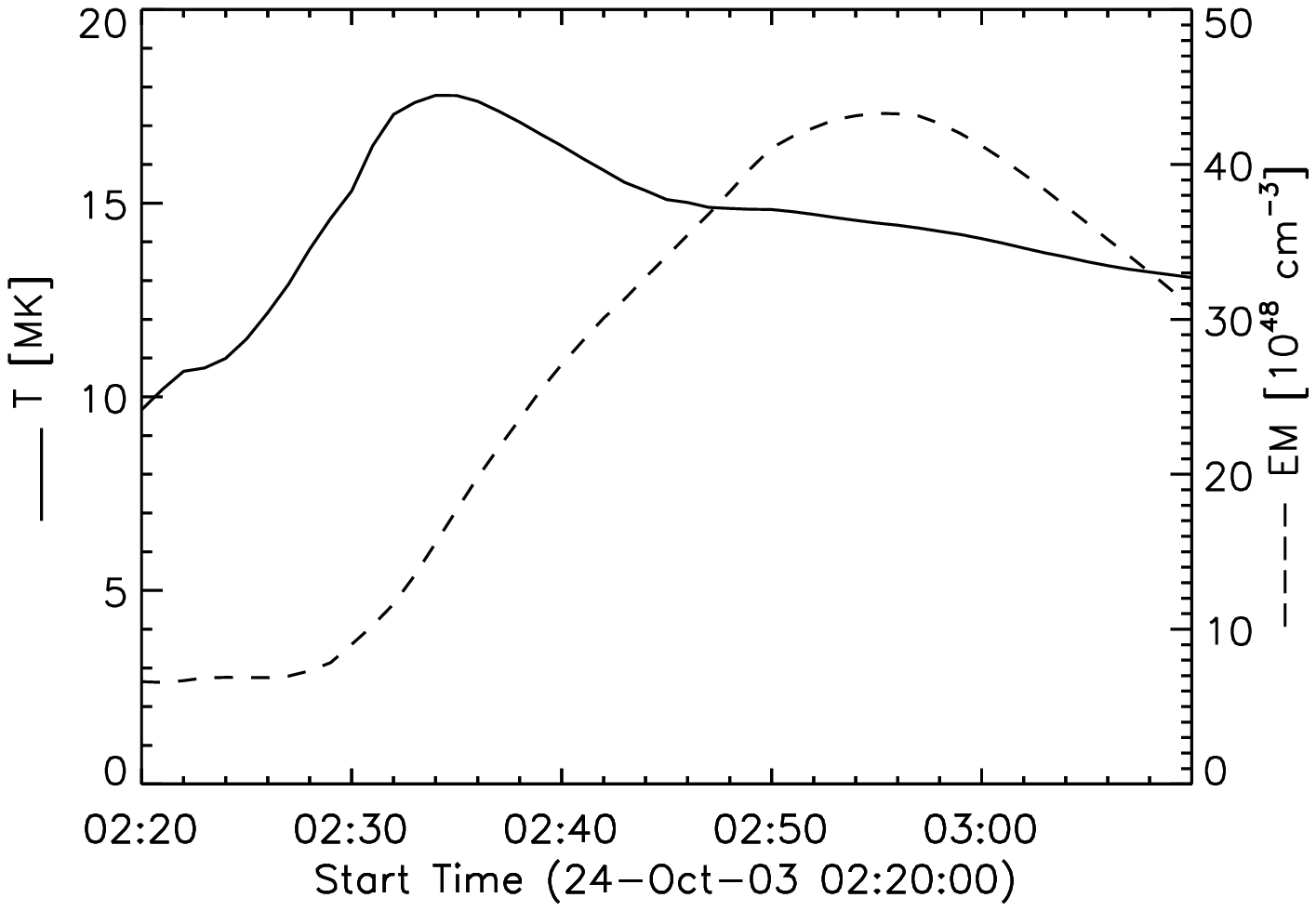}
 \caption{{\em Left:} GOES X-ray fluxes (upper curve: $1-8$ {\AA}, lower curve: $0.5-4$ {\AA}). {\em Right:} Temperature and emission measure during the rise phase of the 2003 October 24 flare obtained from the GOES/XRS data.}
 \label{goes24oct03}
\end{figure}

The flare started at about 02:19 UT in active region NOAA 10486. SXR flux reached its maximum (M7.6) at 02:54 UT. Two phases are clearly seen (see Figure \ref{goes24oct03}, left panel) in the GOES light curves. First one from the beginning up to $\sim$02:44 UT, the second one after 02:44 UT. We used GOES/XRS data to calculate temperature, $T_{G}$, and emission measure, $EM_{G}$, of the whole flare. Temperature increased slowly and reached its maximum value ($17.8$ MK) at 02:34 UT (Figure \ref{goes24oct03}, right panel). Emission measure reached its maximum value ($43\times 10^{48}$~cm$^{-3}$) at 02:55 UT. Difference, $\Delta t$, between these two maxima equals 21 minutes. 

\begin{figure} 
\begin{center} 
  \includegraphics[width=0.58\textwidth,clip=]{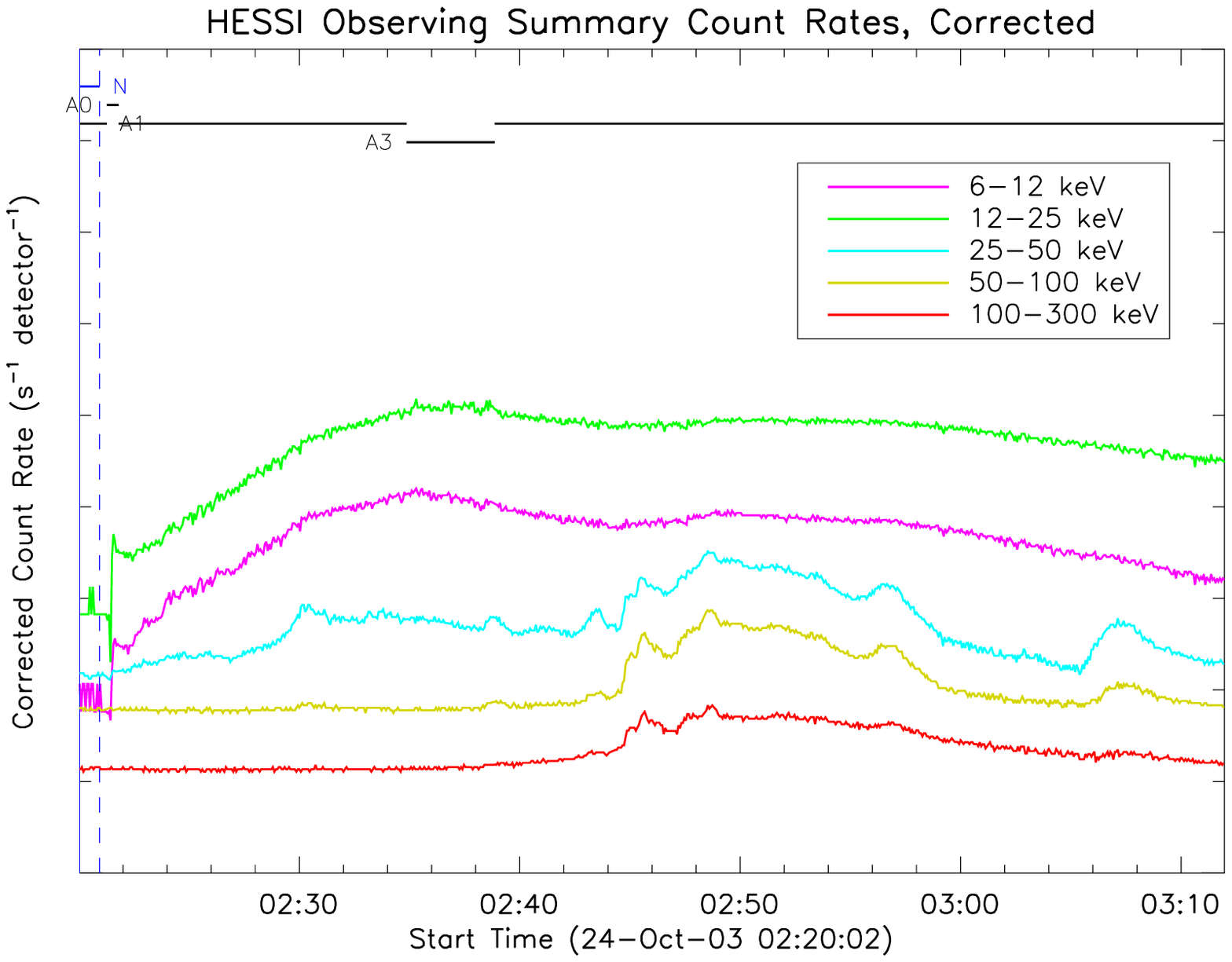}
    \includegraphics[width=0.39\textwidth,clip=]{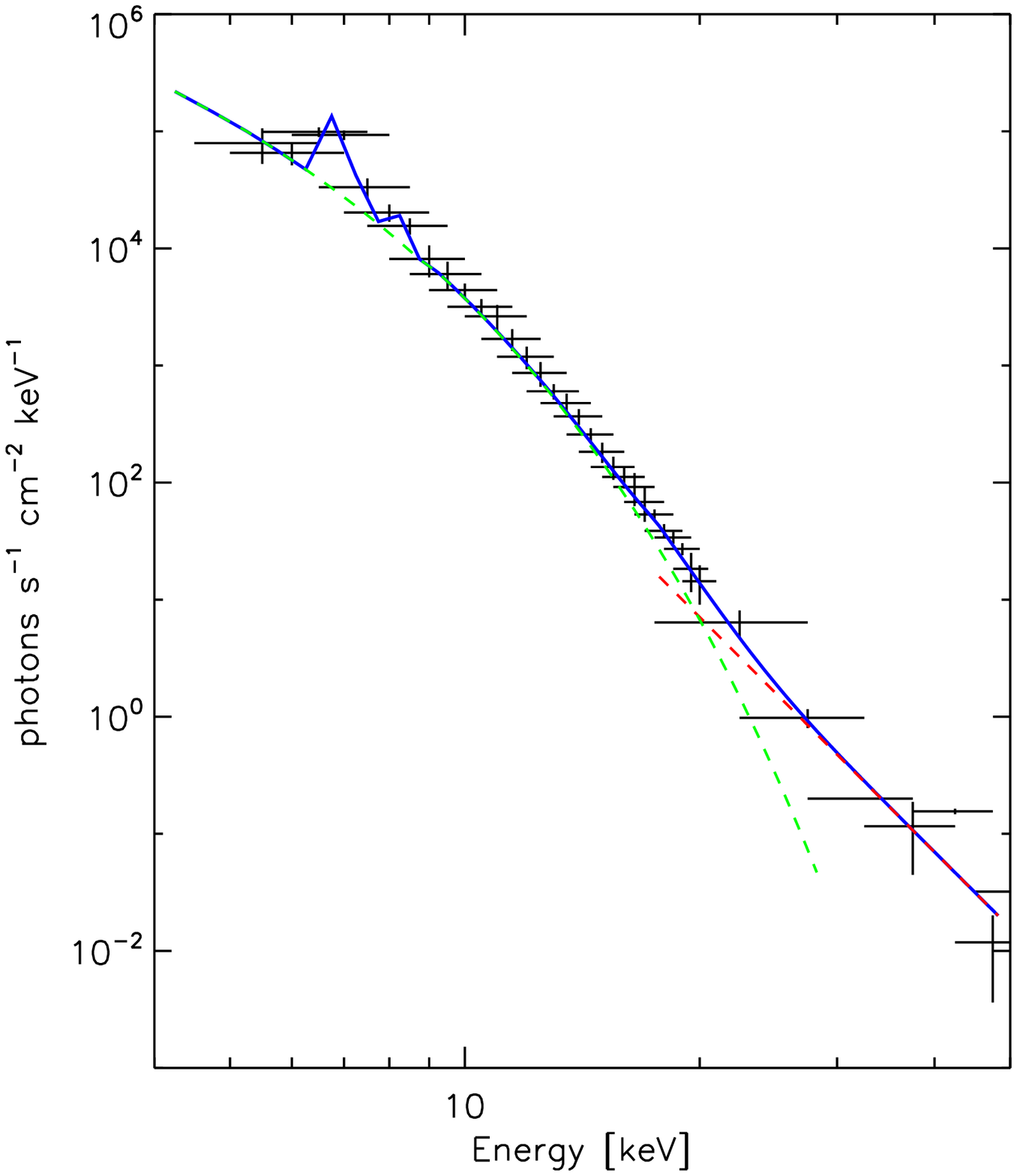}
 \caption{{\em Left:}RHESSI light curves of the 2003 October 24 flare.{\em Right:} The RHESSI X-ray spectrum of the loop-top source (source N) observed during the 2003 October 24 flare at 02:48 UT (horizontal bars corresponds to the energy bin widths). This spectrum was fitted using the thermal component (green curve), two spectral line complexes (at $6.7$~keV and $8.0$~keV) and single power-law component (red curve). The sum of all these models, the best-fit model, is represented by the blue curve.}
 \label{curve24oct03}
\end{center} 
\end{figure}

The flare was well observed by many instruments. Detailed multi-wavelength analysis of this flare was presented by other authors \cite{li2008,joshi2009}. Two phases of this flare are also visible in the RHESSI light curves (Figure \ref{curve24oct03}, left panel).\\

\begin{figure}
 \includegraphics[width=0.48\textwidth,clip=]{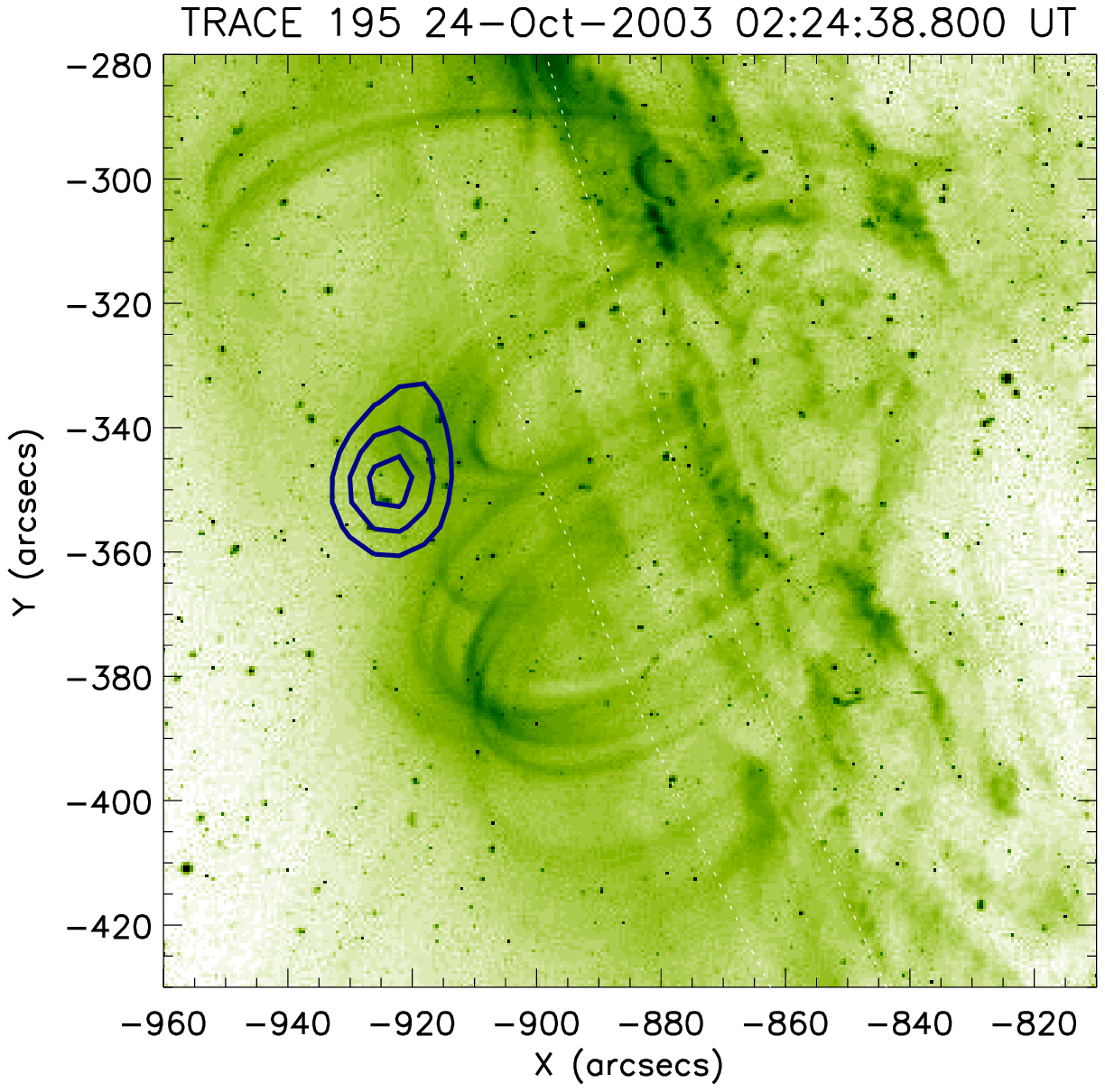}
 \includegraphics[width=0.48\textwidth,clip=]{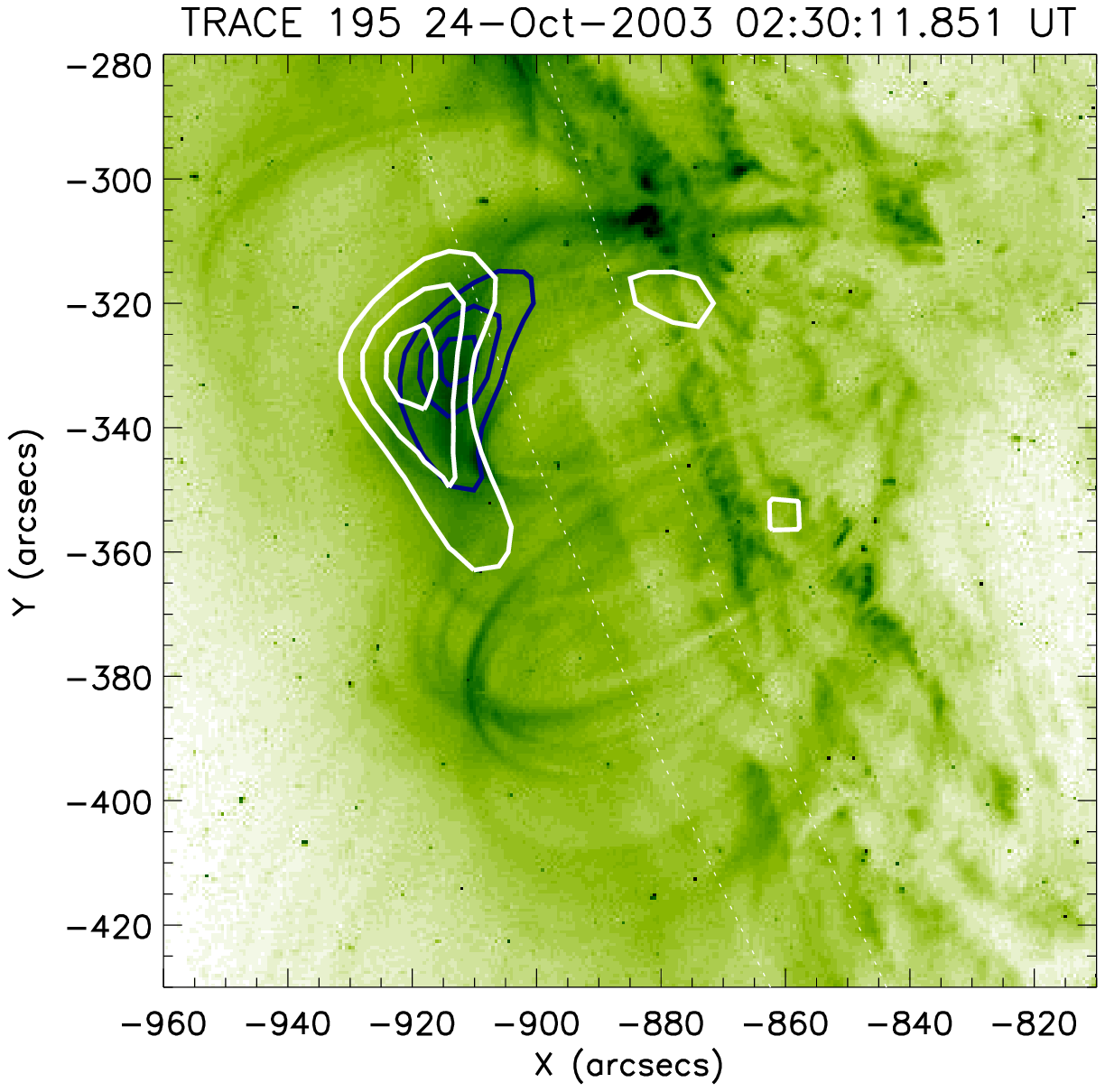}
\caption{TRACE $195$ {\AA} images showing 2003 October 24 flare during the first phase. Contours show the emission in the $10-11$~keV range ({\em blue}) and $20-25$~keV range ({\em white}) observed with RHESSI. The contours are for 50\%, 70\% and 90\% of maximum emission.}
 \label{24oct03_trace195_1f}
 
\end{figure}
\begin{figure}[t] 
 \includegraphics[width=0.48\textwidth,clip=]{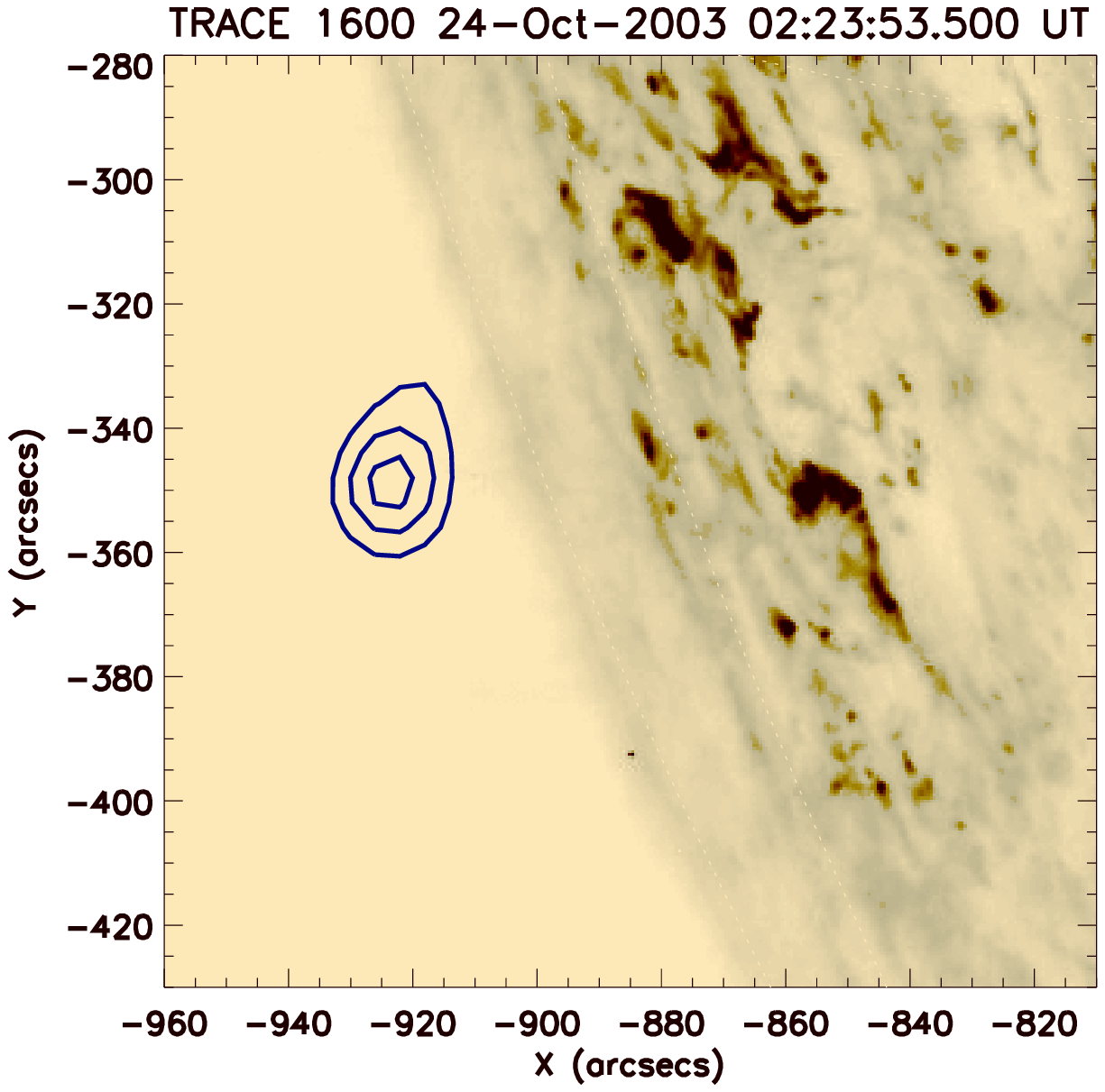}
 \includegraphics[width=0.48\textwidth,clip=]{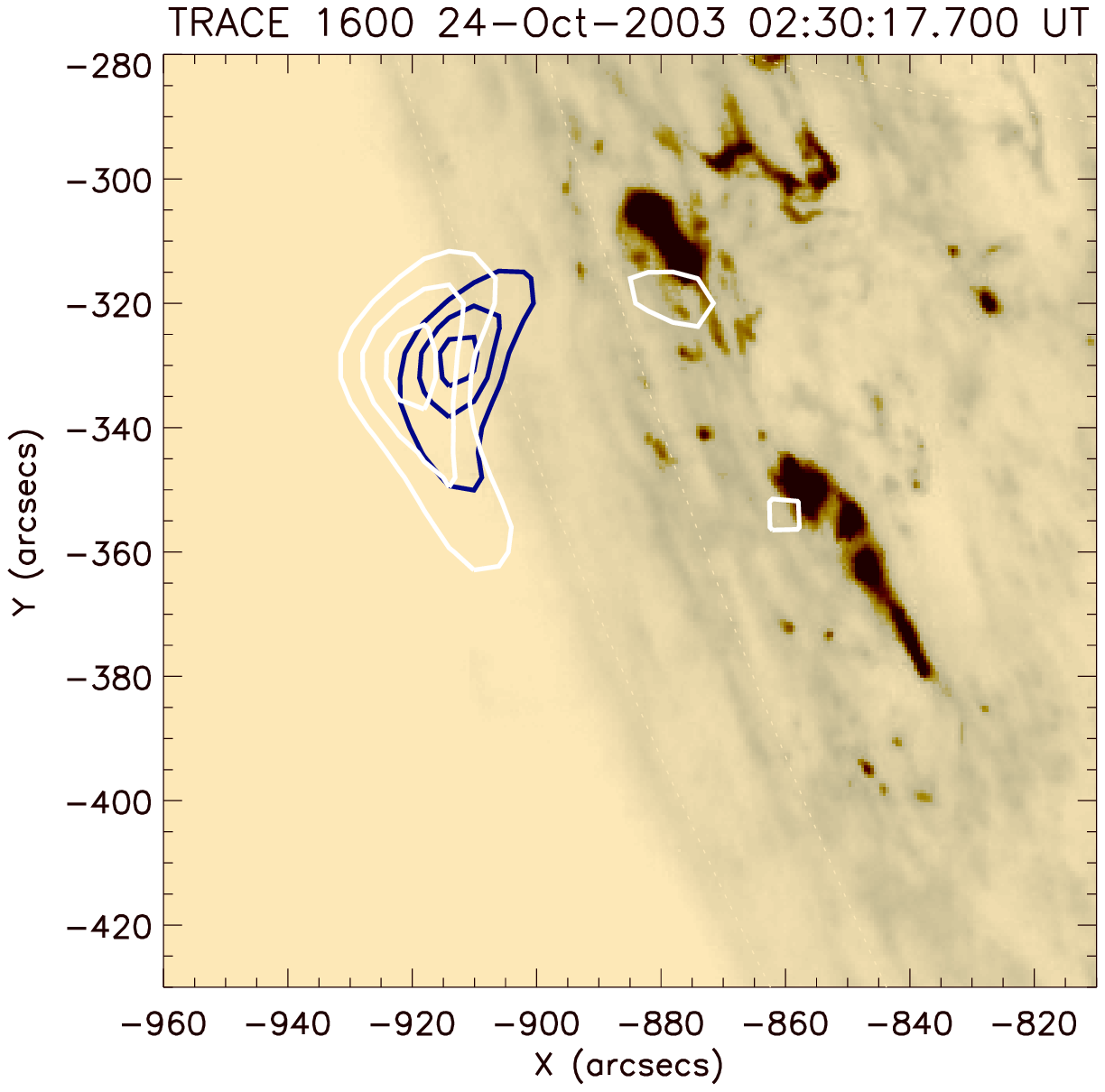}
\caption{TRACE $1600$ {\AA} images showing 2003 October 24 flare during the first phase. Contours show the emission in the $10-11$~keV range ({\em blue}) and $20-25$~keV range({\em white}) observed with RHESSI. The contours are for 50\%, 70\% and 90\% of maximum emission.}
 \label{24oct03_trace1600_1f}
\end{figure}

\subsubsection{First phase}
RHESSI observed this flare from 02:22 UT. During the first phase emission up to $25-50$ keV  range was observed, there were no short-lasting pulses. TRACE data helped us to investigate morphology of the flare. Thermal response function for TRACE $195$ {\AA} filter has two distinct maxima: higher for plasma at $T$~$\sim$~$1-2 \,\rm MK$ and lower for significantly hotter plasma ($T$~$\sim$~$10-30 \,\rm MK$, \opencite{phillips2005a}). We carefully co-alignmented RHESSI and TRACE $195$ {\AA} images using SOHO/EIT $195$ {\AA} images \cite{gallagher2002}. TRACE $195$ {\AA} images with RHESSI contours are shown in Figure~\ref{24oct03_trace195_1f}. 
At the beginning, one source, slightly elongated in the north direction, was observed. A few minutes later, the source (source N) was visible at lower altitude (see Table \ref{table}) and was spatially correlated with the very bright loop-top source seen in the TRACE $195$ {\AA}  images. The hot source N dominated in the emission during the first phase. Shrinkage of TRACE and RHESSI loops was previously reported \cite{li2008,joshi2009}. Only at about 02:30 UT footpoints (FPN -- in north direction, FPS -- in south direction) were visible and were spatially correlated with the ribbons seen in TRACE $1600$ {\AA} images (Figure \ref{24oct03_trace1600_1f}).

\subsubsection{Second phase}
\begin{figure}[!t] 
  \includegraphics[width=0.48\textwidth,clip=]{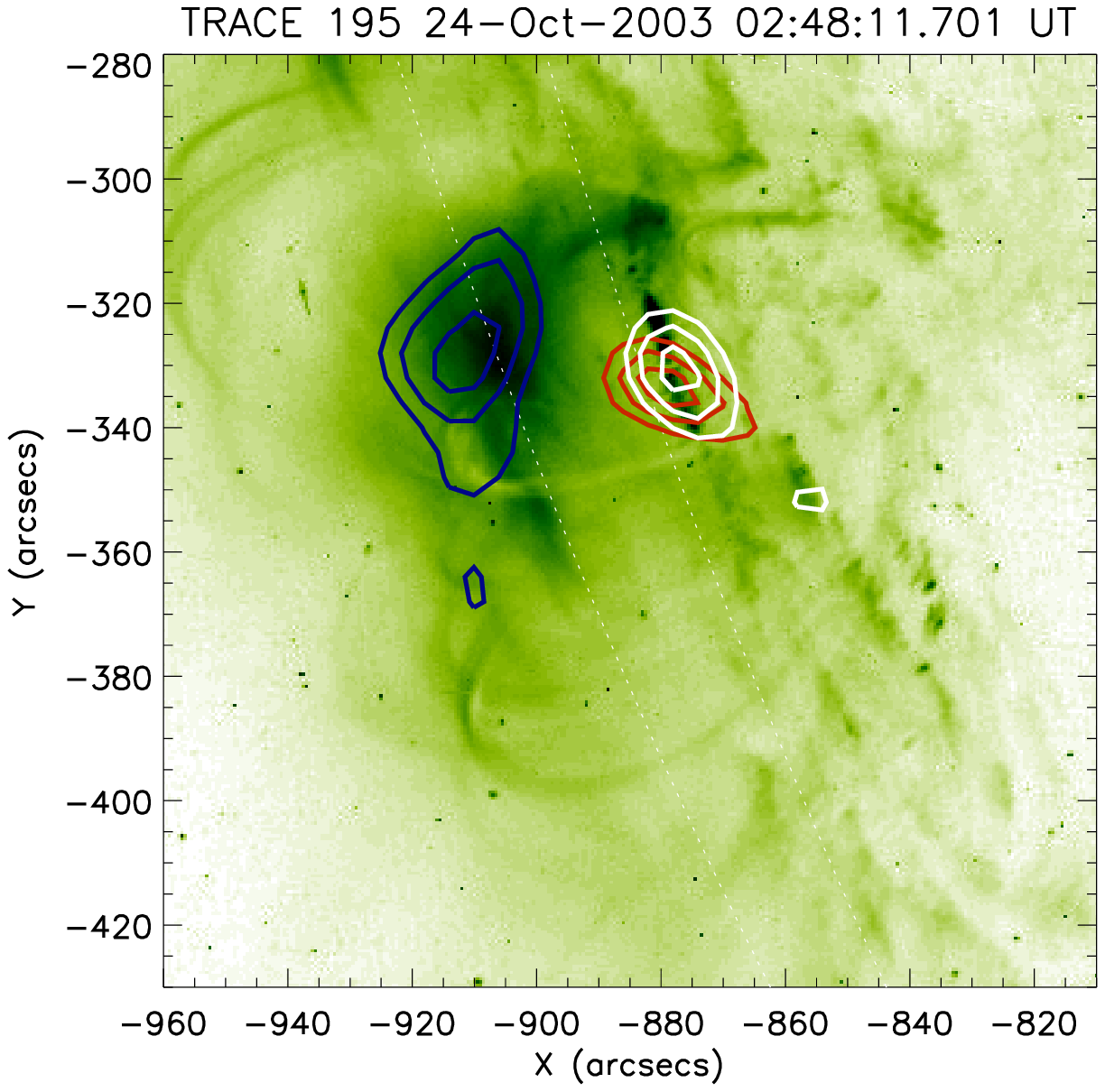}
  \includegraphics[width=0.48\textwidth,clip=]{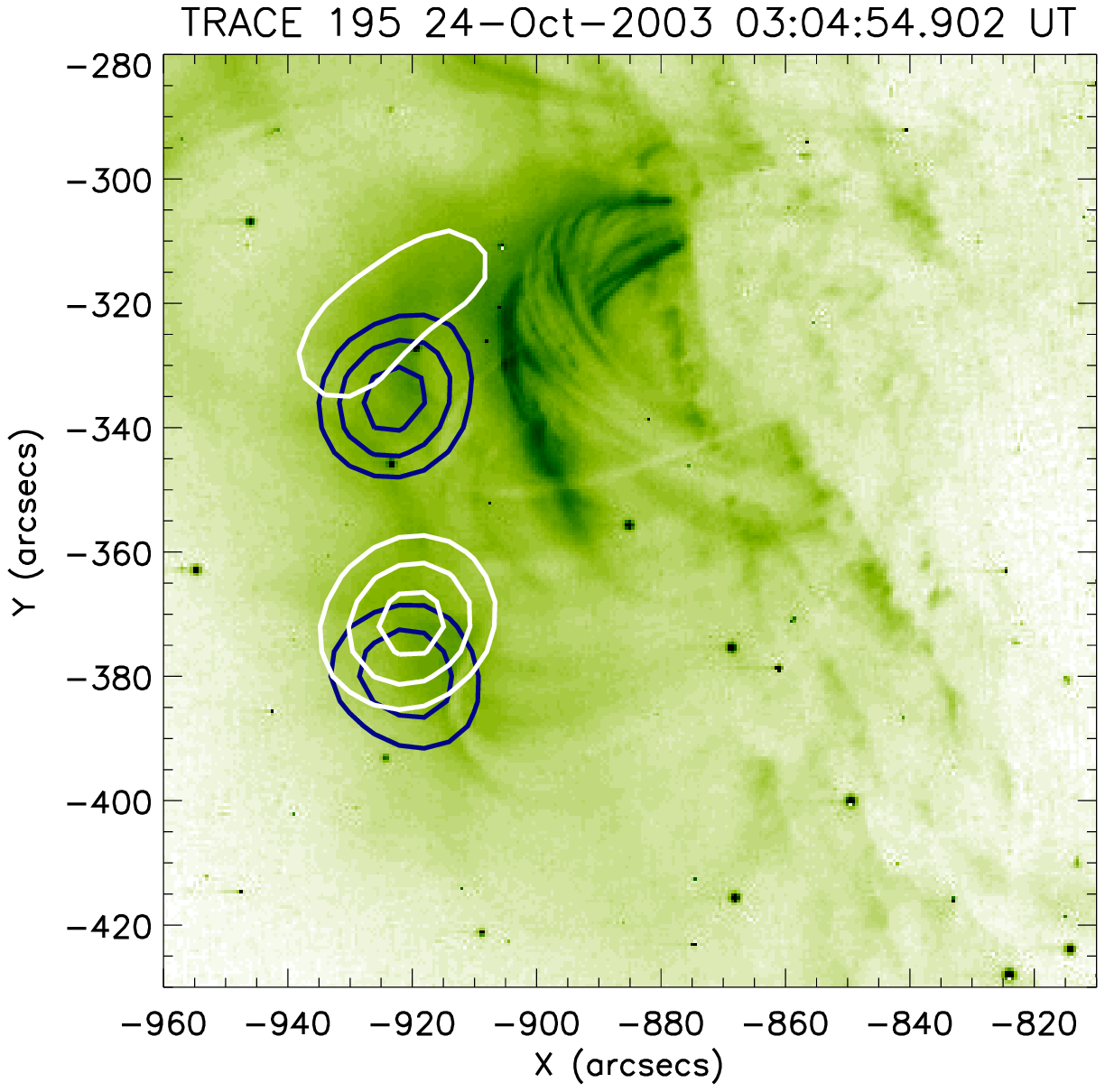}
\caption{TRACE $195$ {\AA} images showing the 2003 October 24 flare during the second phase. Contours show the emission in the $10-11$~keV range ({\em blue}), $20-25$~keV range({\em white}) and $100-120$~keV range ({\em red}) observed with RHESSI. The contours are for 50\%, 70\% and 90\% of maximum emission.}
 \label{24oct03_trace195_2f}
\end{figure}

\begin{figure}[t] 
 \includegraphics[width=0.48\textwidth,clip=]{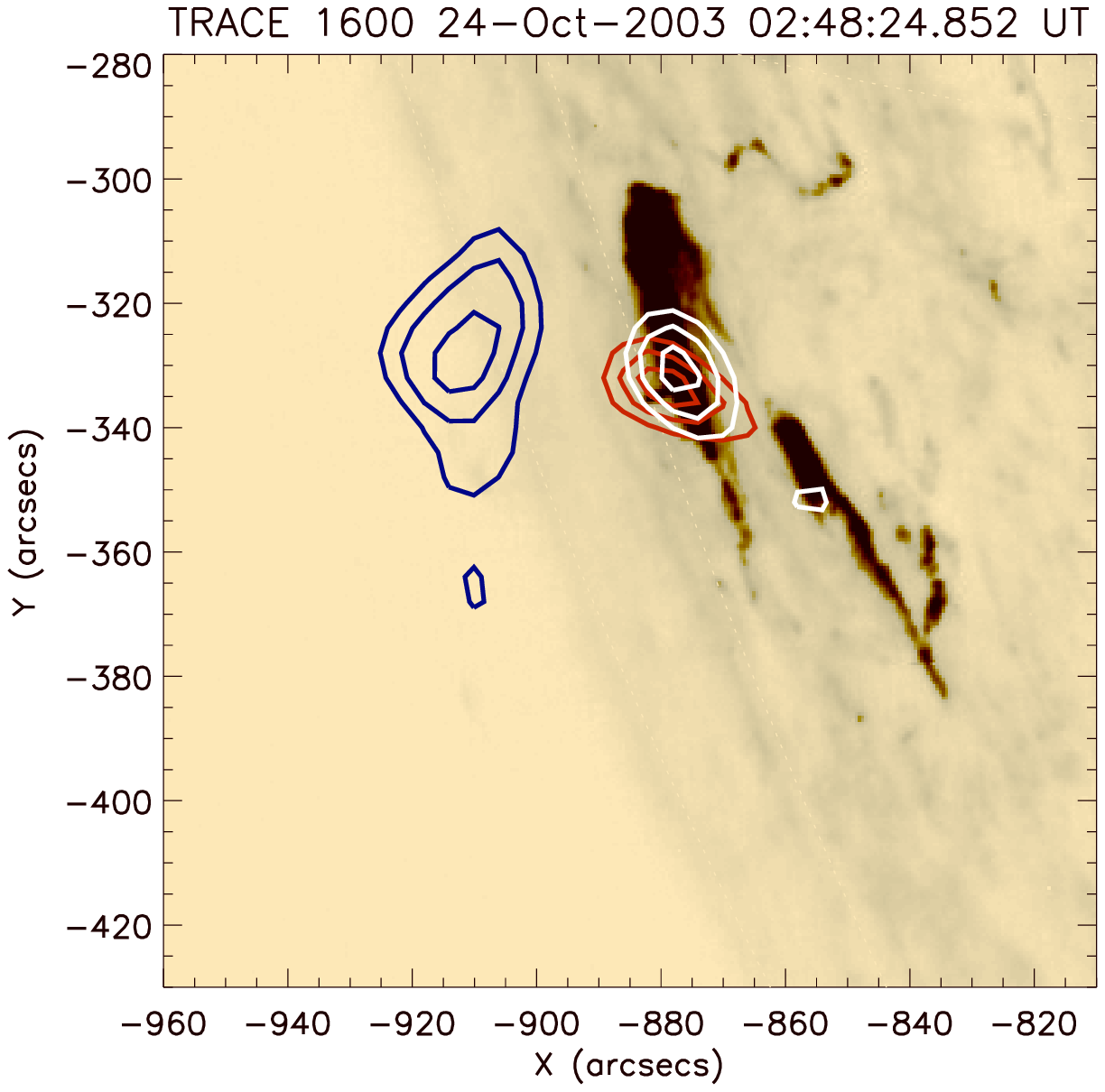}
 \includegraphics[width=0.48\textwidth,clip=]{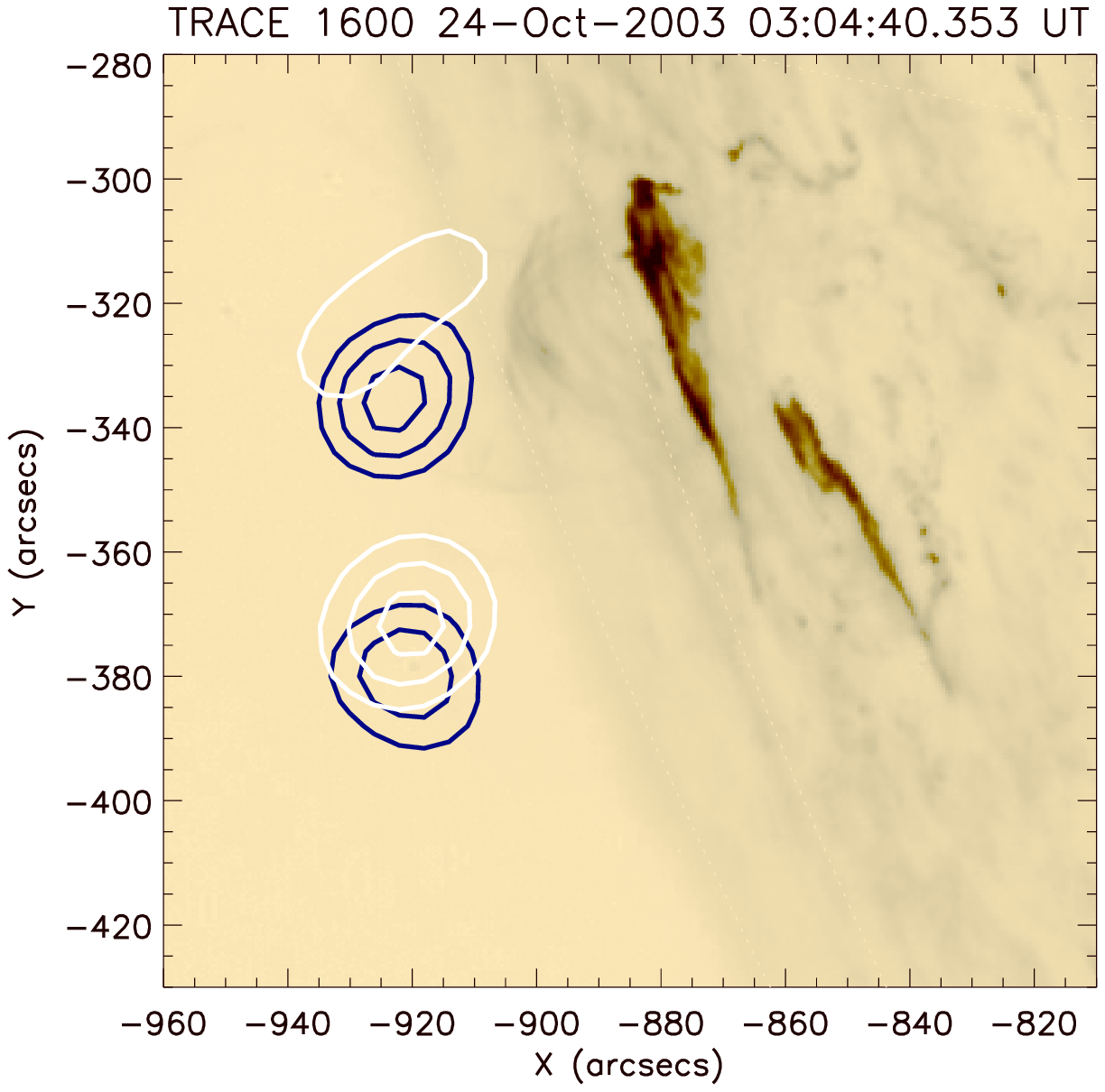}
\caption{TRACE $1600$ {\AA} images showing the 2003 October 24 flare during the second phase. Contours show the emission in the $10-11$~keV range ({\em blue}) and $20-25$~keV range({\em white}) and $100-120$~keV range ({\em red}) observed with RHESSI. The contours are for 50\%, 70\% and 90\% of maximum emission.}
 \label{24oct03_trace1600_2f}
\end{figure}

The second phase started at about 02:44 UT and lasted after flare maximum. Emission in energy range of $100-300$ keV range was observed during this phase. Emission in lower energy range was dominated by source N, which was observed up to $30-35$ keV. At flare maximum second source, S, was observed and brightened with time (Figure \ref{24oct03_trace195_2f}). Several minutes after the maximum of the flare, the source S dominated emission in energy up to $25$ keV. For some time we observed only one elongated source and there was no possibility to separate it into sources N and S.
The emission from footpoints was very strong, especially at the beginning of the second phase (Figure \ref{24oct03_trace1600_2f}). Separation between footpoints was smaller than during the first phase. At the beginning of the second phase FPN was stronger than FPS and was observed in energy range up to $200-300$ keV. FPS was observed in the energy range up to $70-80$ keV after flare maximum (at about 03:07--03:08 UT).

\subsubsection{Physical parameters}
We calculated temperature, $T_{R}$, emission measure, $EM_{R}$, and non-thermal energy parameters of both LTSs from imaging spectroscopy (see example spectrum in Figure \ref{curve24oct03}, right panel). The source N was observed during the whole rising phase and was brighter than source S during most of the time. Plasma in the LTS N was hot at the beginning,  $T_{R}>25$\,MK, and less hot after the flare maximum, $T_{R}\sim16$\,MK, (Table \ref{table}). Heating rate was also very high ($E_{H}>15$\,erg\,cm$^{-3}$s$^{-1}$ at the beginning and $E_{H}<2.5$\,erg\,cm$^{-3}$s$^{-1}$ after the  flare maximum). This value of heating rate is much higher than values obtained from {\em Yohkoh} observations, due to weak {\em Yohkoh}/SXT sensitivity to higher temperature plasma. After reaching its maximum value (at the beginning of the flare), heating rate decreased with characteristic time $\tau=1240$ s. Rate of the non-thermal energy release per unit volume for source N was about $5$\,erg\,cm$^{-3}$s$^{-1}$ during the rising phase.

The source S was observed mostly during the second phase. Maximum value of temperature of LTS S was $\sim21$ MK and decreased very slowly to $\sim18$ MK after flare maximum. After flare maximum the source S remains hotter and brighter than the source N and became dominant source.
 
\subsection{2003 November 18 flare}

\begin{figure}[!b] 
 \includegraphics[width=0.47\textwidth,clip=]{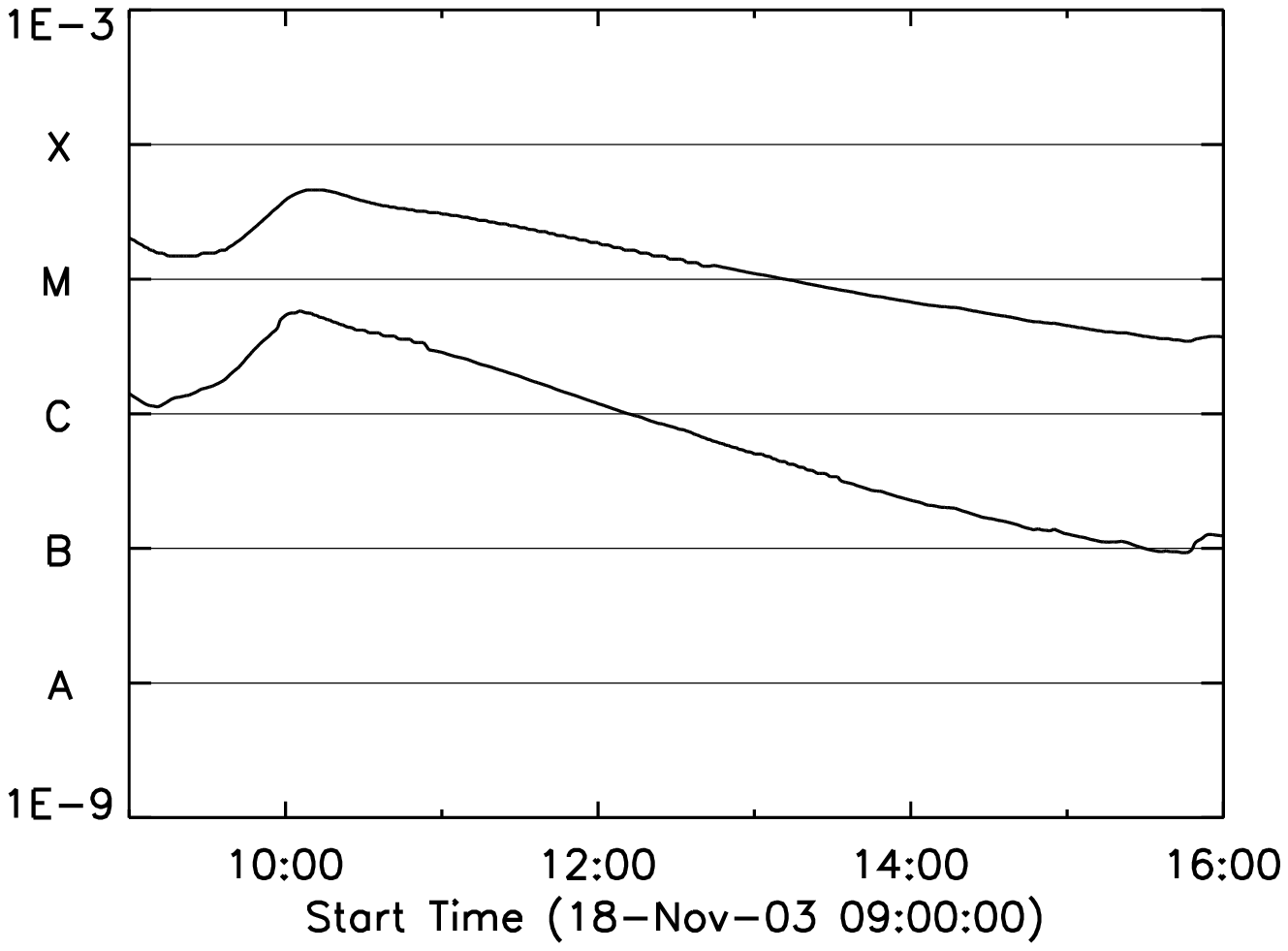}
 \includegraphics[width=0.5\textwidth,clip=]{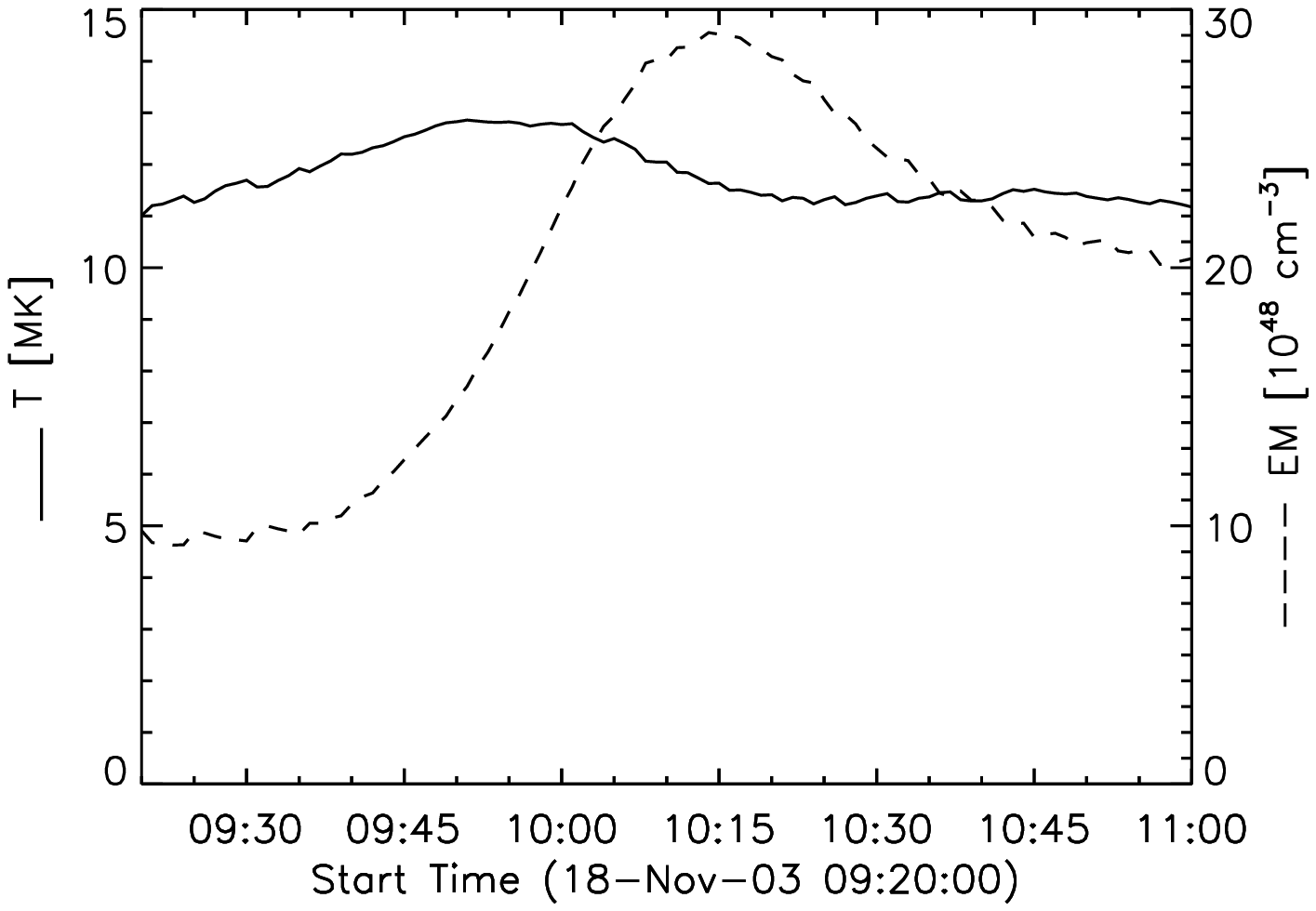}
 \caption{{\em Left:} GOES X-ray fluxes (upper curve: $1-8$ {\AA}, lower curve: $0.5-4$ {\AA}). {\em Right:} Temperature and emission measure during the rise phase of the 2003 November 18 flare obtained from the GOES/XRS data.}
 \label{goes18nov03}
\end{figure}

The flare occurred in active region NOAA 10506 at the eastern solar limb. It  began at 09:23 UT. SXR flux increased slowly and reached its maximum value (GOES class M4.5) at 10:11 UT (Figure {\ref{goes18nov03}}, left panel). GOES fluxes were used to calculate temperature, $T_{G}$, and emission measure, $EM_{G}$, and time shift between their maxima. Temperature increased slowly and reached its maximum value ($12.8$ MK) at 09:51 UT (Figure {\ref{goes18nov03}}, right panel). After that, temperature slightly dropped (down to $11$ MK). After 10:30 UT, $T_{G}$ slighty increased which was related to additional emission, seen on GOES light curve after the maximum. Emission measure reached its maximum value ($29\times 10^{48}$~cm$^{-3}$) at 10:14 UT. Time difference between $T_{G}$ and $EM_{G}$ maxima is equal to $23$ minutes.

\begin{figure} 
\begin{center} 
 \includegraphics[width=0.58\textwidth,clip=]{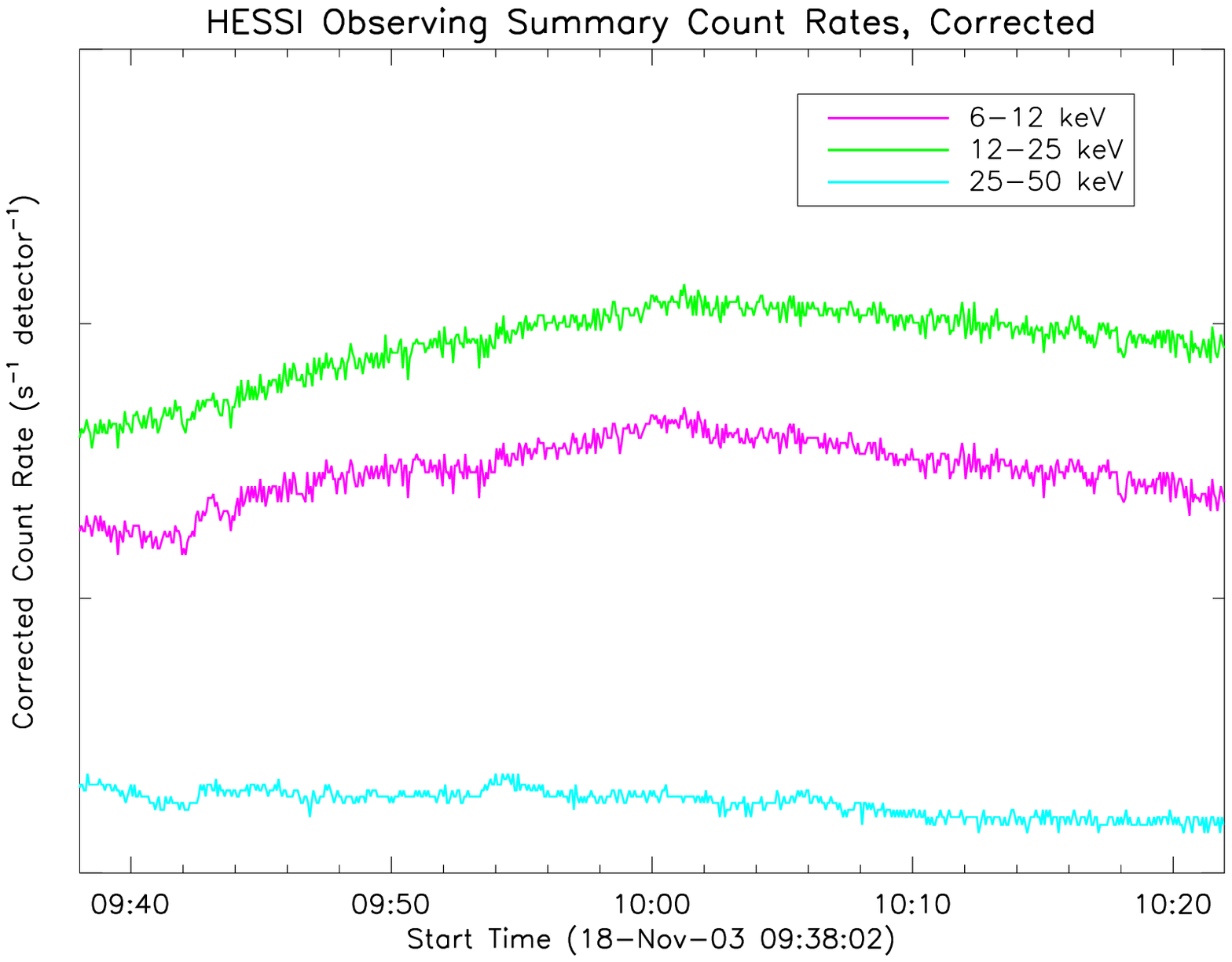}
  \includegraphics[width=0.39\textwidth,clip=]{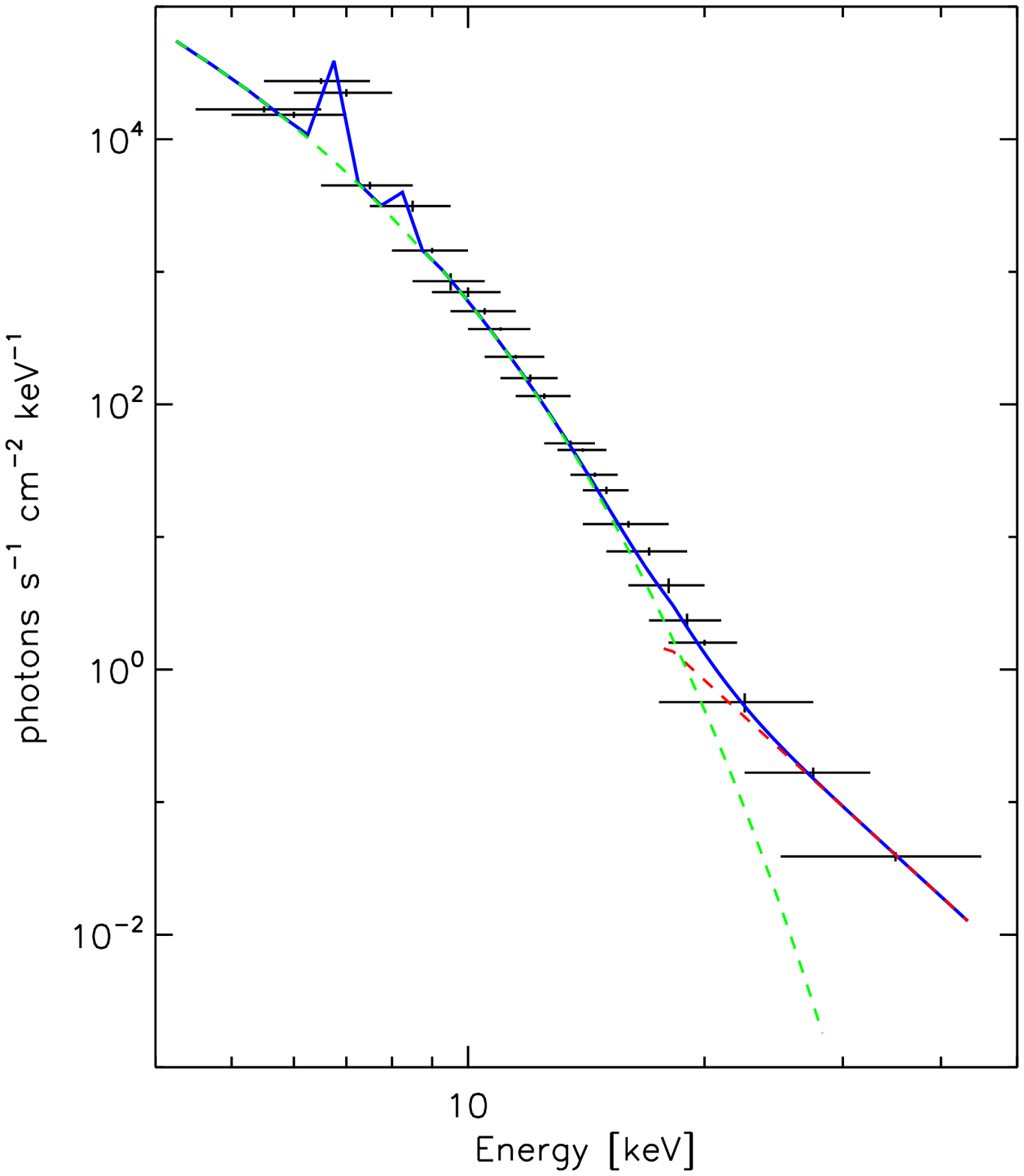}
 \caption{{\em Left:}RHESSI light curves for the 2003 November 18 flare.{\em Right:} The RHESSI X-ray spectrum of the loop-top source observed during the 2003 November 18 flare at 09:46 UT (horizontal bars corresponds to the energy bin widths). This spectrum was fitted using the thermal component (green curve), two spectral line complexes (at $6.7$~keV and $8.0$~keV) and single power-law component (red curve). The sum of all these models, the best-fit model, is represented by the blue curve.}
 \label{curve18nov03}
\end{center} 
\end{figure}

RHESSI observed the flare during almost entire rise phase (see Figure {\ref{curve18nov03}}). Short-lasting pulses were not observed. Only long-lasting emission in energy range up to $25$ keV was observed most of the time. However, for a few moments of time very weak emission in energy range $25-50$ keV was also observed. 

\begin{figure} 
 \includegraphics[width=0.48\textwidth,clip=]{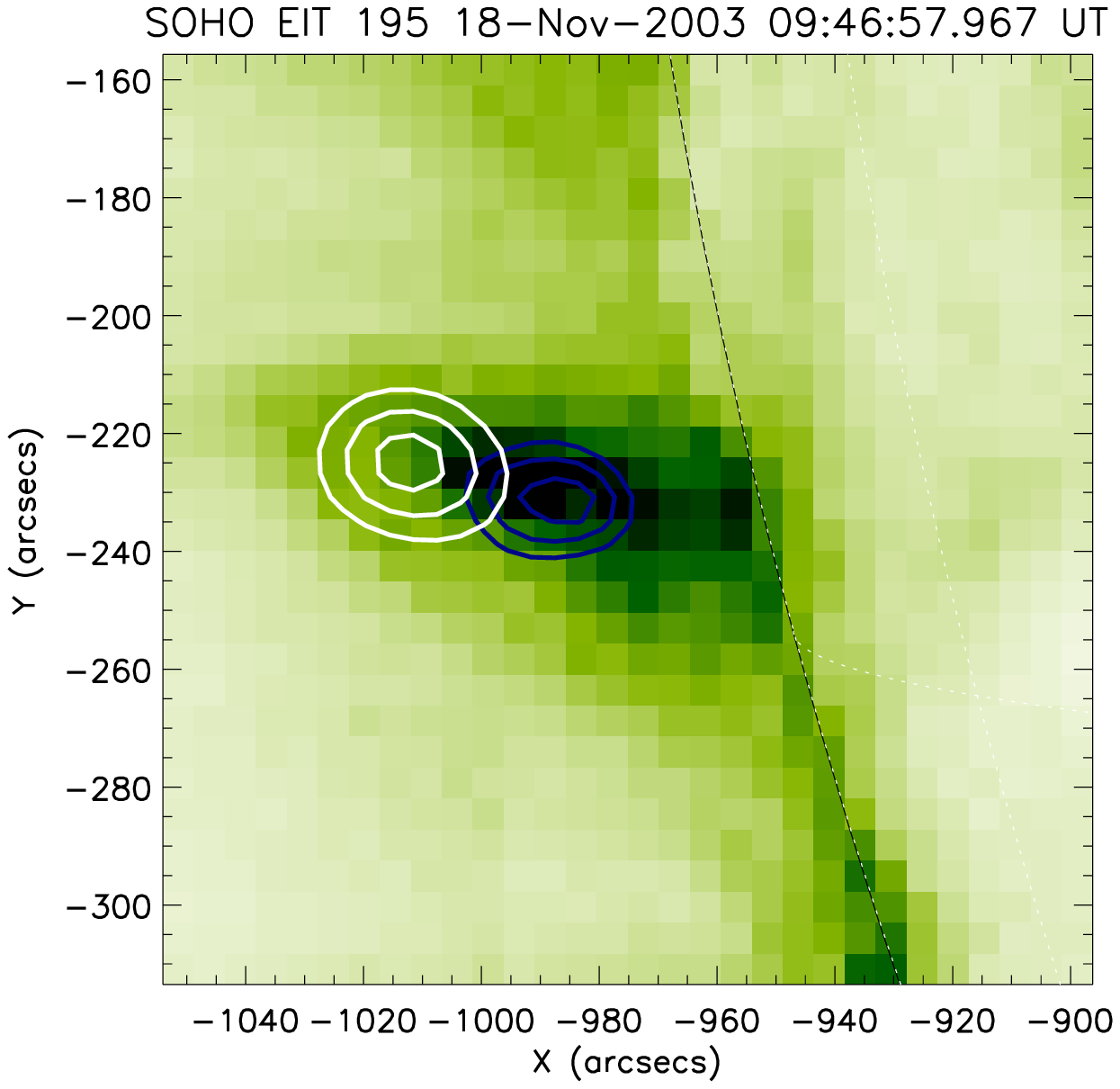}
 \includegraphics[width=0.48\textwidth,clip=]{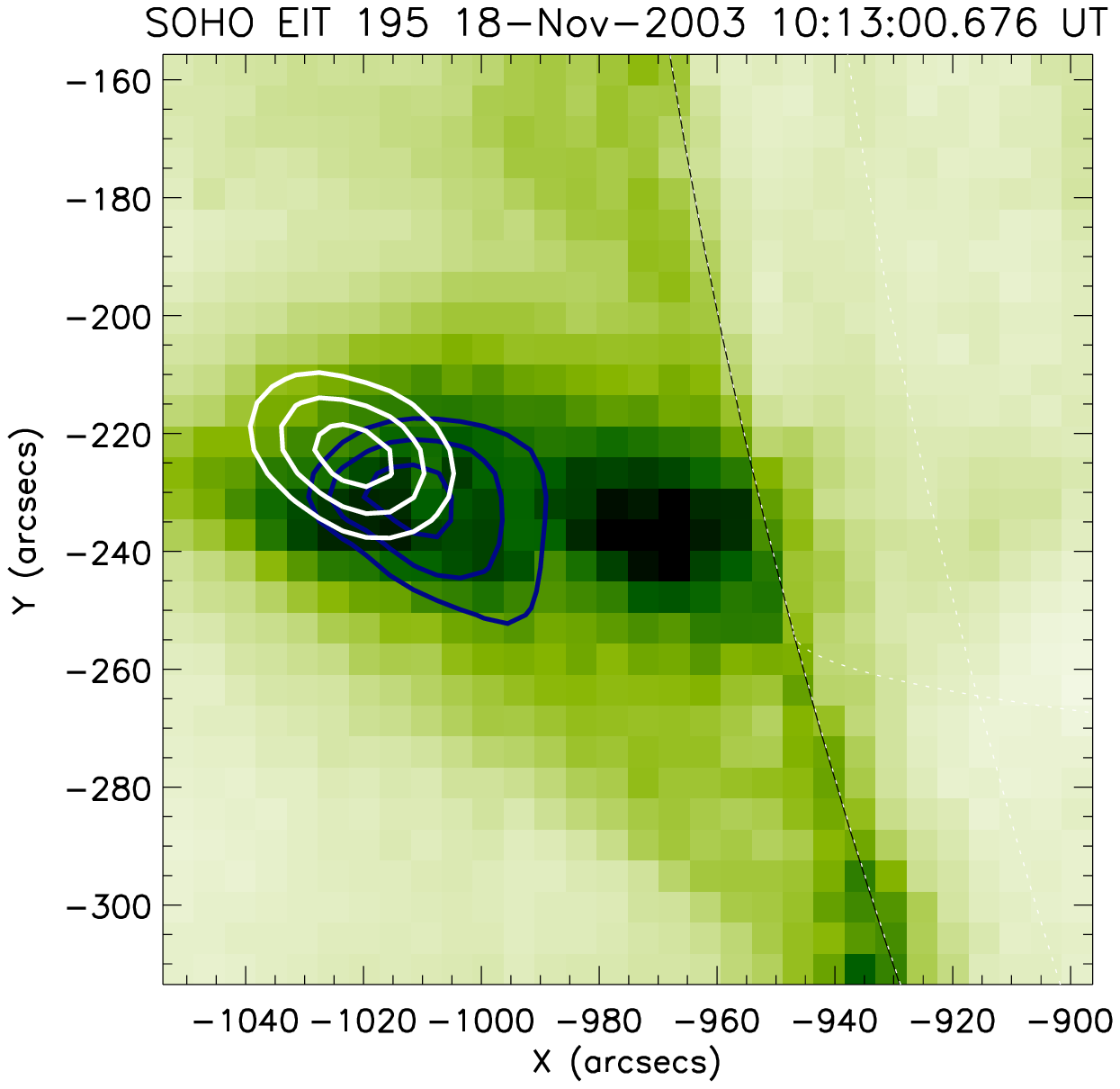}
 \caption{SOHO/EIT $195$ {\AA} images showing flare of the 2003 November 18 during the rise phase. Contours show the emission in the $10-11$~keV range ({\em blue}) and $20-25$~keV ({\em white}) observed with RHESSI. The contours are for 50\%, 70\% and 90\% of maximum emission.}
 \label{obrazki18nov03}
\end{figure}

In the PIXON images only a loop-top source was observed. This source was spatially correlated with the limb structure seen in the SOHO/EIT images (Figure \ref{obrazki18nov03}). Only footpoints of the loops were occulted. The whole structure became visible in the EIT images a few hours after flare maximum, due to the Sun rotation. The structure expanded and a few hours after flare maximum high arcade of loops was observed. We obtained the location of the HXR source, its size and height above the photosphere from PIXON images. Emission at higher energy range ($> 20$ keV) was observed at higher altitude (Figure \ref{obrazki18nov03}). Therefore we calculated average value of height at each moment of time. The average height increased during the rise phase from $h\sim20$\,Mm to $h\sim50$\,Mm. We calculated temperature, $T_{R}$, emission measure, $EM_{R}$, and non-thermal component parameters from the spectra fitting (see Figure \ref{curve18nov03}, right panel). Temperature decreased slowly ($\sim19$\,MK at the beginning, $\sim17$\,MK at maximum), which is characteristic feature of SLDE flares. Physical parameters obtained (see Table \ref{table}) allowed us to calculate thermal energy release rate ($E_{H}\sim5$\,erg\,cm$^{-3}$s$^{-1}$ at the beginning, $\sim1$\,erg\,cm$^{-3}$s$^{-1}$ at the maximum). 

During the rise phase loop-top source was visible up to $40$ keV and its emission showed clear non-thermal component. At about 09:54--09:55 UT rate of the non-thermal energy release per unit volume was about $0.3$ erg cm~$^{-3}$~s$^{-1}$.
 
\subsection{2005 September 6 flare}
\begin{figure}[b] 
 \includegraphics[width=0.449\textwidth,clip=]{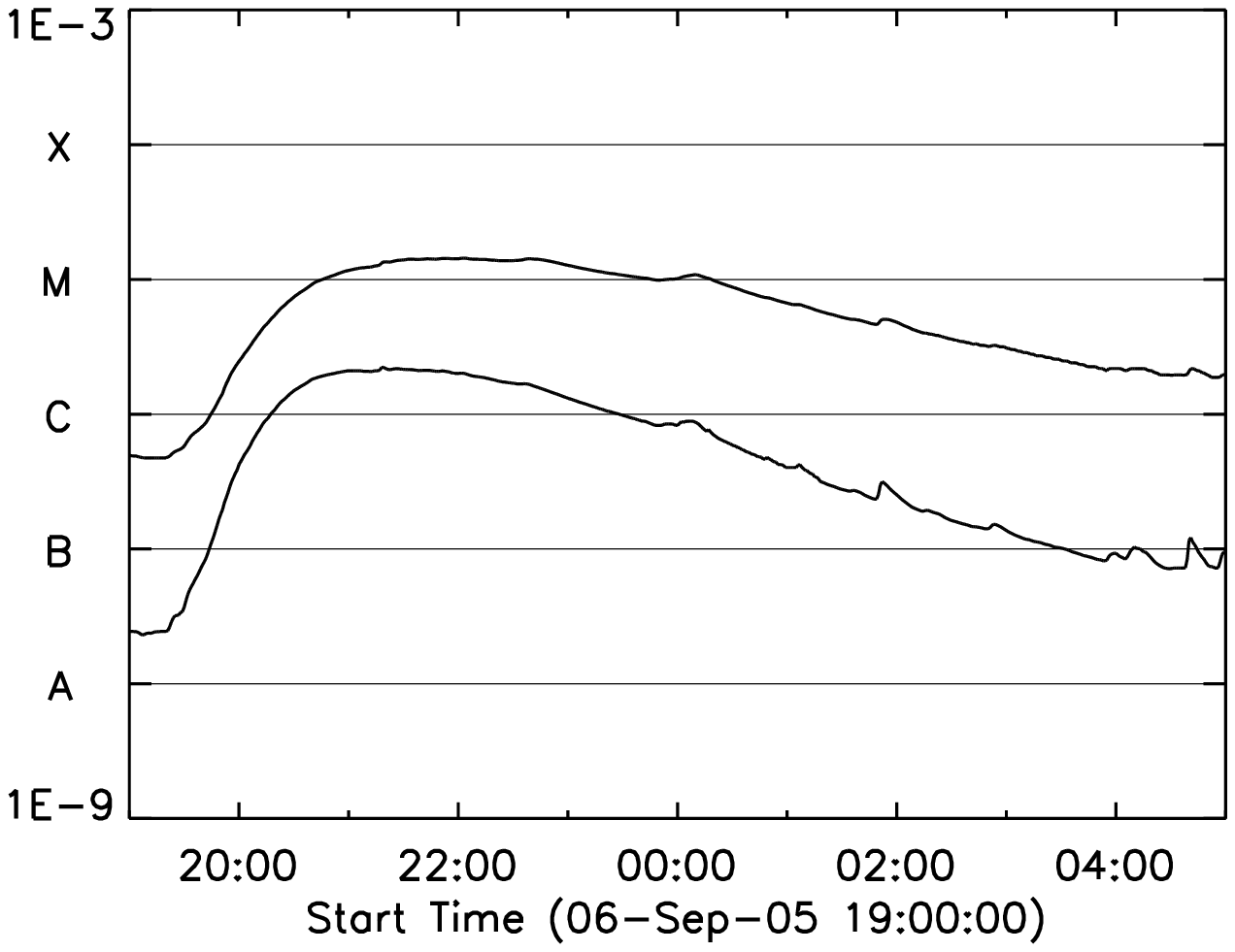}
 \includegraphics[width=0.5\textwidth,clip=]{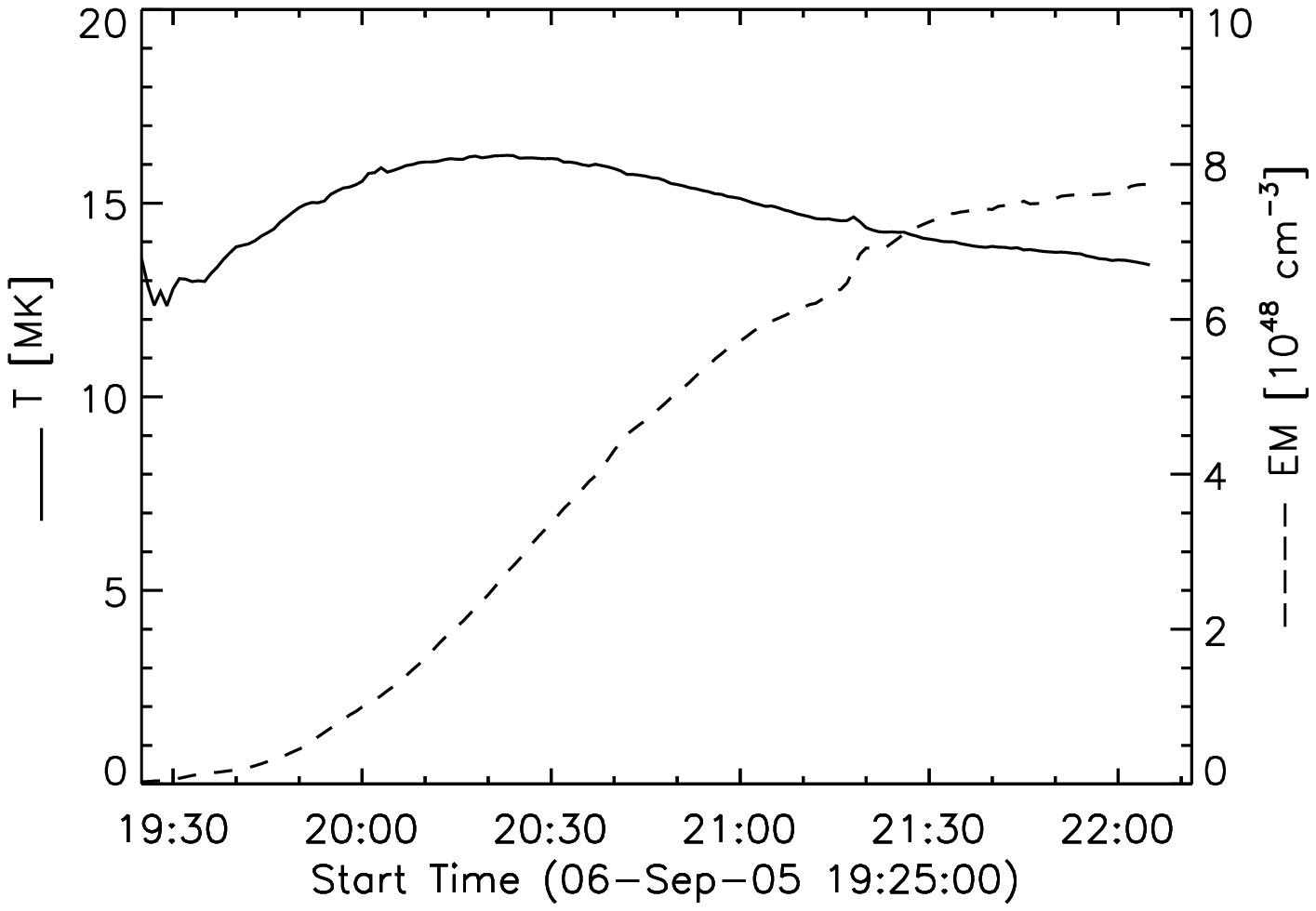}
 \caption{{\em Left:} GOES X-ray fluxes (upper curve: $1-8$ {\AA}, lower curve: $0.5-4$ {\AA}). {\em Right:} Temperature and emission measure during the rise phase of the 2005 September 6 flare obtained from the GOES data.}
 \label{goes6sep05}
\end{figure}
The flare of 2005 September 6 occurred in an active region NOAA 10808, $\sim 6^\circ$ behind the eastern solar limb. The flare (GOES class M1.4) started at 19:32 UT and reached its maximum at 22:02 UT (Figure {\ref{goes6sep05}}, left panel). Temperature obtained from the GOES/XRS data increased for almost $50$ minutes and reached maximum ($16.2$ MK) at 20:23 UT. Decrease of temperature during the rise phase of the flare was very slow (see Figure \ref{goes6sep05}, right panel). Maximum value of emission measure ($7.7\times 10^{48}$~cm$^{-3}$) was at 22:02 UT. Time difference between these two maxima is equal to $101$ minutes, which is the largest $\Delta t$ we have found.\\

\begin{figure} 
\begin{center} 
 \includegraphics[width=0.7\textwidth,clip=]{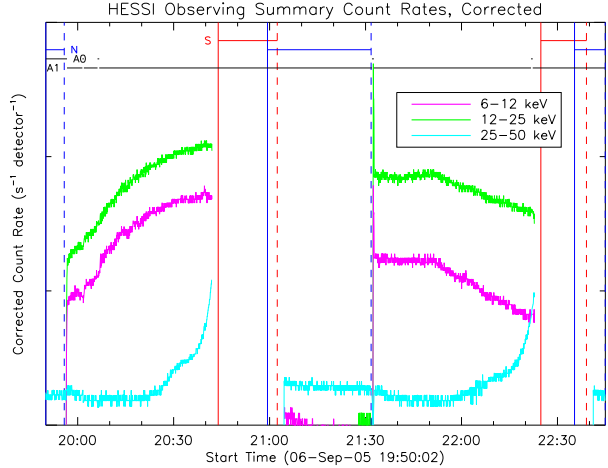}
 \caption{RHESSI light curves for the September 6, 2005 flare.}
 \label{curve6sep05}
\end{center} 
\end{figure} 

\begin{figure} 
 \includegraphics[width=0.5\textwidth,clip=]{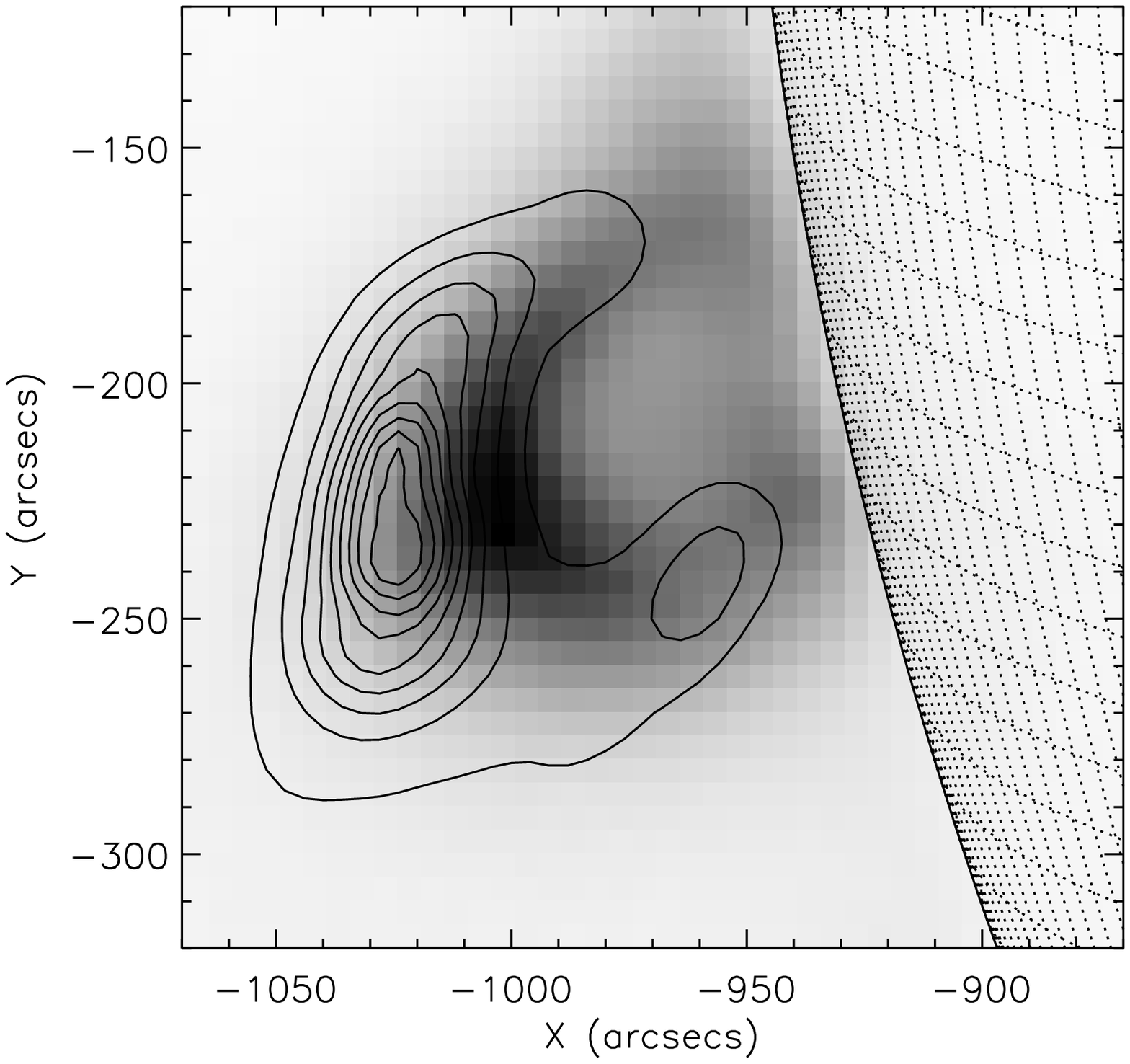}
 \includegraphics[width=0.458\textwidth,clip=]{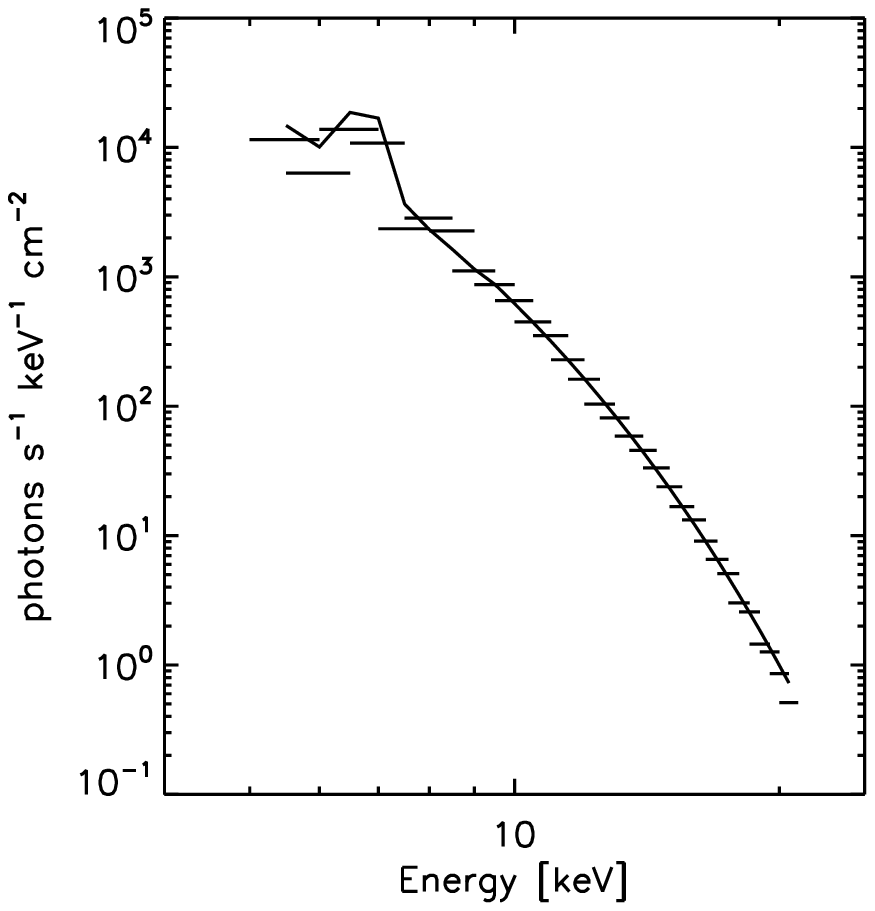}
 \caption{{\em Left:} GOES/SXI image showing 2005 September 6 flare during the rise phase. Contours show the emission in the $8-9$~keV range observed with RHESSI. The contours are for 10\%, 20\%, 30\%, 40\%, 50\%, 60\%, 70\%, 80\% and 90\% of maximum emission. {\em Right:} The RHESSI X-ray spectrum of the loop-top source observed during the 2005 September 6 flare at 20:37 UT (horizontal bars corresponds to the energy bin widths). This spectrum was fitted using the thermal component and two spectral line complexes (at $6.7$~keV and $8.0$~keV). The sum of all these models, the best-fit model, is represented by the black curve.}
 \label{rhessi}
\end{figure}

During the rise phase long-lasting HXR emission up to $25$ keV was observed (Figure {\ref{curve6sep05}}) without short-lasting pulses. This flare was slightly occulted, only footpoints were behind the limb. The HXR emission source is visible above a loop seen in the SXI image (Figure {\ref{rhessi}, left panel}). TRACE observations ($3$ hours after flare maximum) show arcade of loops located near the limb, with loops oriented along the line of sight, which explains "cusp-like" shape seen in SXI images. From the RHESSI PIXON images we determined size of the LTS ($\sim 16-20$ Mm) and its height above the photosphere ($> 70$ Mm). The results are shown in Table \ref{table}. We performed imaging spectroscopy to obtain physical parameters of the LTS. A spectrum with fitted model (thermal component plus lines at $6$~keV and $8$~keV) is shown in Figure {\ref{rhessi}, right panel}. Temperatures obtained ($\sim 20$~MK at the beginning and $\sim 15$~MK at the maximum) are higher than those obtained from the GOES/XRS data which is due to a different temperature sensitivity of these two instruments. Slow changes of temperature are common feature for GOES and RHESSI observations of this flare. Obtained temperature and emission measure (from spectra fitting), size of LTS and height above the photosphere (from images) allowed us to calculate energy balance. Value of heating rate obtained is low (Table \ref{table}) and decreases with the characteristic time of 4430 s, which explains very long rising phase.

\begin{table}[!ht]
	\caption{Physical parameters of LTS obtained from the RHESSI data}
	\vspace{3mm}
	\label{table}
		\begin{tabular}{|c|c|c|c|c|c|c|c|c|}
\hline

&Date&&&&&&&\\
&GOES Start&time&$h$&$r$&$T_{R}$&$EM_{R}$*&$E_{H}$**&$\tau$\\
&GOES Max&&\tiny{(Mm)}&\tiny{(Mm)}&\tiny{(MK)}&&&\tiny{(s)}\\
\hline
&&\multicolumn{7}{c|}{Loop-top source N}\\
&&02:27&34&11.1&25.2&0.4&9.6&\\
&&02:32&27&8.8&25.6&2.4&15.3&1240\\
&&02:48&25&8.9&19.4&6.9&6.3&\\
1.&2003 Oct. 24&03:04&42&8.2&16.5&5.4&2.2&\\
&02:19&\multicolumn{7}{c|}{Loop-top source S}\\
&02:54&02:54&47&4.0&20.8&1.6&9.3&\\
&&03:02&46&7.6&21.0&1.3&5.1&\\
&&03:10&50&6.3&18.6&1.9&3.6&\\
\hline
2.&2003 Nov. 18&09:40&21&12.0&19.2&0.6&5.1&\\
&09:23&09:50&30&11.4&17.7&2.4&2.8&2110\\
&10:11&10:01&33&13.9&17.6&3.9&2.1&\\
&&10:10&40&12.2&17.4&3.9&2.0&\\
\hline
3.&2005 Jul. 13&14:16&38&8.5&23.5&2.3&7.4&\\
&14:01&14:23&41&9.0&22.1&4.7&5.3&2000\\
&14:49&14:32&39&9.3&22.0&7.1&5.3&\\
&&14:39&40&9.8&20.1&10.0&3.6&\\
\hline
4.&2005 Aug. 23&14:20&27&9.2&19.7&0.04&5.9&\\
&14:19&14:25&25&7.8&24.3&0.5&14.5&650\\
&14:44&14:29&25&7.6&21.4&2.7&11.0&\\
&&14:33&28&7.4&20.6&3.8&7.8&\\
\hline
&&19:58&72&19.8&20.3&0.2&1.0&\\
5.&6-Sep-05&20:17&72&16.4&20.8&0.4&1.3&\\
&19:32&20:37&77&17.7&20.4&0.8&1.1&4430\\
&22:02&21:34&84&19.0&16.8&1.5&0.5&\\
&&21:49&84&19.4&16.2&1.5&0.4&\\
&&22:05&82&20.3&15.2&1.9&0.3&\\
\hline
6.&2007 Jan. 25&06:47&27&12.8&15.8&0.1&1.6&\\
&06:33&06:55&29&12.4&13.7&1.5&1.0&1100\\
&07:14&07:00&29&14.0&12.9&2.4&0.7&\\
\hline

\multicolumn{9}{@{} l @{}}{* $EM$ in $10^{48}$cm$^{-3}$} \\
\multicolumn{9}{@{} l @{}}{** $E_{H}$ in erg\,cm$^{-3}$s$^{-1}$} \\
	\end{tabular}
\end{table}

\section{Summary}
Slow Long Duration Events are flares characterized by long rising phase and smooth HXR emission. We used RHESSI and GOES/XRS data to obtain physical parameters of loop-top sources of such flares. TRACE, SOHO/EIT and GOES/SXI images helped us to investigate morphology of the flares. 

Our analysis can be summarized as follows:
\begin{itemize} 
\item Using RHESSI and GOES/XRS observations, we have confirmed that characteristic feature of the SLDEs is a large time interval, $\Delta t>20$ minutes, between maximum of temperature and maximum of emission measure. 
\item As other LDEs, slow LDEs often occur in arcade of loops. Height of these structures are high ($h\sim25-50$ Mm). 
\item In almost all cases (except flare No. 4) long-lasting HXR emission was observed. Flare with the longest rise phase (No. 5) was $\sim 6^\circ$ behind the limb so we did not observe footpoints. In four cases (No. 1, 2, 3, 4), non-thermal emission from the loop-top sources was also observed. 
\item In flares No. 3 and 4 rate of non-thermal electron energy release per unit volume (from LTS) was comparable to the heating rate. 
\item For all analysed flares we calculated value of thermal energy release rate. Obtained values of $E_H$ are larger than those previous obtained from SXT data (\opencite{bak2005}). This difference is caused by the fact that SXT telescope had limited sensitivity to higher temperature plasma ($>$~$10$~MK) which usually led to underestimation of flare temperature. RHESSI data enable us to estimate much more reliable (higher) values of plasma temperature and therefore higher values of $E_H$. 
\item We calculated characteristic time $\tau$ of heating function decrease after reaching its maximum value. Long duration of the SLDEs rising phase is consistent with a very slow decrease of the $E_{H}$ during that phase. In most cases characteristic time of $E_{H}$ decrease is larger than $1000$~s. We obtained the highest value of $\tau$ for the flare with the longest rising phase (No. 5). We obtained the lowest value of $\tau$ for the flare with the shortest rise phase (No. 4).
\item Assuming $E_{H}$ equal to minimum value of $E_{H}$ calculated for the rising phase, average volume of LTS and duration od rising phase obtained from GOES/XRS data, we estimated total thermal energy released during the rising phase. In our flares total thermal energy is at least $10^{31}$ ergs. The same magnitude of energy is released during the decay phase of LDEs \cite{kolomanski2011}. This value is larger than value of total thermal energy released during rise phase of short-rising flares (\opencite{jiang2006}). Those authors obtained value not higher than $10^{30}$ ergs.
\end{itemize} 

\section*{Acknowledgements} 
The RHESSI satellite is NASA Small Explorer (SMEX) mission. We thank Professor Jerzy Jakimiec for many inspiring discussions and also thank Barbara Cader-Sroka for editorial remarks. We thank anonymous referee for useful comments and suggestions. This investigation has been supported by a Polish Ministry of Science and High Education, grant No. N203 1937 33.

%
%

\begin{thebibliography}{}

\bibitem[\protect\citeauthoryear{Aschwanden}{2005}]{aschwanden2005}
Aschwanden, M. J.: 2005, {\em Physics of the Solar Corona}: An Introduction, Praxis Publishing (Chichester, UK) and Springer (Berlin).
 
\bibitem[\protect\citeauthoryear{B\c ak-St\c e\' slicka and Jakimiec}{2005}]{bak2005}
B\c ak-St\c e\' slicka, U., Jakimiec, J.: 2005, {\em Solar Phys}. \textbf{231}, 95.

\bibitem[\protect\citeauthoryear{B\c ak-St\c e\' slicka}{2007}]{bak2007}
B\c ak-St\c e\' slicka, U.: 2007, {\em PhD Thesis}, University of Wroc{\l aw}

\bibitem[\protect\citeauthoryear{Delaboudini`ere {\em et al.}}{1995}]{delaboudiniere1995}
Delaboudini`ere, J.-P., Artzner, G. E., Brunaud, J., Gabriel, A. H., Hochedez, J. F. et al.: 1995, {\em Solar Phys}. \textbf{162}, 19.

\bibitem[\protect\citeauthoryear{Dere {\em et al.}}{2009}]{dere2009}	                                                                             
Dere, K. P., Landi, E., Young P. R., Del Zanna, G., Landini, M. et al.: 2009, {\em A$\& $A} \textbf{498}, 915.

\bibitem[\protect\citeauthoryear{Donnelly, Grubb, and Cowley }{1977}]{donnelly77}	        
Donnelly, R. F., Grubb, R. N., Cowley, F. C.: 1977, NOAA Tech. Memo. ERL SEL-48.

\bibitem[\protect\citeauthoryear{Feldman {\em et al.}}{1995}]{feldman1995}
Feldman, U., Seely, J. F., Doschek, G. A., Brown, C. M., Phillips, K. J. H. et al.: 1995, {\em Astrophys. J.} \textbf{446}, 860.

\bibitem[\protect\citeauthoryear{Gallagher {\em et al.}}{2002}]{gallagher2002}
Gallagher, P. T., Dennis, B. R., Krucker, S., Schwartz, R. A., Tolbert, A. K.: 2002, {\em Solar Phys.} \textbf{210}, 341.

\bibitem[\protect\citeauthoryear{Handy, Acton, and Kankelborg}{1999}]{handy1999}
Handy, B.N., Acton, L.W., Kankelborg, C.C.: 1999, {\em Solar Phys.} \textbf{187}, 229.

\bibitem[\protect\citeauthoryear{Harra-Murnion, Schmieder, and van Driel-Gesztelyi}{1998}]{harra1998}
Harra-Murnion, L. K., Schmieder, B., van Driel-Gesztelyi, L.: 1998, {\em A$\& $A} \textbf{337}, 911.

\bibitem[\protect\citeauthoryear{Hill {\em et al.}}{2005}]{hill2005}
Hill, S. M., Pizzo, V. J., Balch, C. C., Biesecker, D. A., Bornmann, P. et al.: 2005, {\em Solar Phys.} \textbf{226}, 255.


\bibitem[\protect\citeauthoryear{Hudson, Acton, and Freeland}{1996}]{hudson1996}
Hudson, H. S., Acton, L. W., Freeland, S. L.: 1996, {\em Astrophys. J.} \textbf{470}, 629.

\bibitem[\protect\citeauthoryear{Hudson, Canfield, and McKenzie}{1978}]{hudson1978}
Hudson, H.S., Canfield, R.C., and Kane, S.R.: 1978, {\em Solar Phys.} \textbf{60}, 137.
				
\bibitem[\protect\citeauthoryear{Hudson and McKenzie}{2000}]{hudson2000}
Hudson, H. S., McKenzie, D. E.: 2000,  High Energy Solar Physics Workshop - Anticipating HESSI, ASP Conference Series, Vol. 206. Edited by R. Ramaty and N. Mandzhavidze, 221.

\bibitem[\protect\citeauthoryear{Hudson and McKenzie}{2001}]{hudson2001}
Hudson, H. S., McKenzie, D. E.: 2001, {\em Earth Planets Space} \textbf{53}, 581.

\bibitem[\protect\citeauthoryear{Hurford, Schmahl, and Schwartz}{2002}]{hurford2002}
Hurford, G. J., Schmahl, E. J., Schwartz, R. H.: 2002, {\em Solar Phys}. \textbf{210}, 61.

\bibitem[\protect\citeauthoryear{Isobe {\em et al.}}{2002}]{isobe2002}
Isobe, H., Yokoyama, T., Shimojo, M., Morimoto, T., Kozu, H. et al.: 2002, {\em Astrophys. J.} \textbf{566}, 528.

\bibitem[\protect\citeauthoryear{Jakimiec {\em et al.}}{1992}]{jakimiec92}
Jakimiec, J., Sylwester, B., Sylwester, J., Serio, S., Peres, G. et al.: 1992, {\em A$\& $A} \textbf {253}, 269.

\bibitem [\protect\citeauthoryear{Jakimiec {\em et al.}}{1997}]{jakimiec1997} 
Jakimiec, J., Tomczak, M., Fludra, A., Falewicz, R.: 1997, {\em Adv. in Space Res.} \textbf{20}, 2341.

\bibitem [\protect\citeauthoryear{Jiang {\em et al.}}{2006}]{jiang2006} 
Jiang, Y. W., Liu, S., Liu, W., Petrosian, V.: 2006, {\em Astrophys. J.} \textbf{638}, 1140.

\bibitem[\protect\citeauthoryear{Joshi {\em et al.}}{2009}]{joshi2009}
Joshi, B., Veronig, A., Cho, K.-S., Bong, S. C., Somov, B. V et al.: 2009, {\em Astrophys. J.} \textbf{706}, 1438.

\bibitem[\protect\citeauthoryear{Kahler}{1977}]{kahler1977}
Kahler, S.: 1977, {\em Astrophys. J.} \textbf{214}, 891.

\bibitem[\protect\citeauthoryear{Ko\l oma\'nski}{2007a}]{kolomanski2007a}
Ko\l oma\'nski, S.: 2007a, {\em A$\& $A} \textbf{465}, 1021.

\bibitem[\protect\citeauthoryear{Ko\l oma\'nski}{2007b}]{kolomanski2007b}
Ko\l oma\'nski, S.: 2007b, {\em A$\& $A} \textbf{465}, 1035.

\bibitem[\protect\citeauthoryear{Ko\l oma\'nski, Mrozek, and B\c ak-St\c e\' slicka}{2011}]{kolomanski2011}
Ko\l oma\'nski, S., Mrozek, T., B\c ak-St\c e\' slicka, U.: 2011, submitted to {\em A$\& $A}.

\bibitem[\protect\citeauthoryear{Kosugi {\em et al.}}{1991}]{kosugi1991}
Kosugi, T., Masuda, S., Makishima, K., Inda, M., Murakami, T. et al.: 1991, {\em Solar Phys}. \textbf{136}, 17.

\bibitem [\protect\citeauthoryear{Li and Li}{2008}]{li2008} 
Li, H., Li, Y.: 2008, {\em Adv. in Space Res.} \textbf{41}, 962.

\bibitem [\protect\citeauthoryear{Lin {\em et al.}}{2002}]{lin2002} 
Lin, R.P., Dennis, B.R., Hurford, G.J., Smith, D.M., Zehnder, A., et al.: 2002, {\em Solar Phys}. \textbf{210}, 3.

\bibitem[\protect\citeauthoryear{McKenzie and Hudson}{1999}]{mckenzie1999}
McKenzie, D. E., Hudson, H., S.: 1999, {\em Astrophys. J.} \textbf{519}, L93.

\bibitem[\protect\citeauthoryear{McKenzie}{2000}]{mckenzie2000}
McKenzie, D. E.: 2000, {\em Solar Phys}. \textbf{195}, 38.

\bibitem[\protect\citeauthoryear{Phillips, Chifor, and Landi}{2005a}]{phillips2005a}
Phillips, K. J. H., Chifor, C., Landi, E.: 2005, {\em Astrophys. J.} \textbf{626}, 1110.

\bibitem[\protect\citeauthoryear{Phillips, Feldman, and Harra}{2005b}]{phillips2005b}
Phillips, K. J. H., Feldman, U., Harra, L. K.: 2005, {\em Astrophys. J.} \textbf{634}, 641.

\bibitem [\protect\citeauthoryear{Pi\~na and Puetter}{1993}]{pina93}
Pi\~na, R.K., Puetter, R.C.: 1993, {\em PASP} \textbf{105}, 630.

\bibitem[\protect\citeauthoryear{Puetter and Yahil}{1999}]{puetter1999} 
Puetter, R.C., Yahil, A.: 1999, Astronomical Data Analysis Software and Systems VIII, ASP Conference Series 172, 307.

\bibitem [\protect\citeauthoryear{Reeves and Warren}{2002}]{reeves2002}
Reeves, K.K., Warren, H.P.: 2002, {\em Astrophys. J.} \textbf{578}, 590.

\bibitem[\protect\citeauthoryear{Sheeley {\em et al.}}{1975}]{sheeley1975} 
Sheeley, N. R., Jr., Bohlin, J. D., Brueckner, G. E., Purcell, J. D., Scherrer, V. E. et al.: 1975, {\em Solar Phys}. \textbf{45}, 377.

\bibitem[\protect\citeauthoryear{Tomczak}{1997}]{tomczak1997}
Tomczak, M.: 1997, {\em A$\& $A} \textbf{317}, 223.

\bibitem[\protect\citeauthoryear{Tsuneta {\em et al.}}{1991}]{tsuneta1991}
Tsuneta, S., Acton, L., Bruner, M., Lemen, J., Brown, W. et al.: 1991, {\em Solar Phys.} \textbf{136}, 37.

\bibitem[\protect\citeauthoryear{Tsuneta {\em et al.}}{1992}]{tsuneta1992}
Tsuneta, S., Hara, H., Shimizu, T., Acton, L. W., Strong, K. T. et al.: 1992, {\em PASJ} \textbf{44}, 63.

\end{thebibliography}
%

\end{article} 
\end{document}